\documentclass[aps,prd,reprint, superscriptaddress,preprintnumbers,nofootinbib]{revtex4-2}
\usepackage{amsmath,amssymb, amsthm,amstext}
\usepackage{natbib}
\usepackage{graphicx}
\usepackage{color}
\usepackage{array, enumerate}
\usepackage{bm}
\usepackage{multirow}
\usepackage[breaklinks,colorlinks,citecolor=blue]{hyperref}
\usepackage{braket}
\usepackage{txfonts}
\usepackage[dvipsnames]{xcolor}        
\usepackage{tabularx}

\begin{document}
\preprint{LIGO-P2300223}
\title{Probing  Spin-Induced Quadrupole Moments in Precessing Compact Binaries }
\author{Zhenwei Lyu}
\email{zwlyu@pku.edu.cn}
\affiliation{Kavli Institute for Astronomy and Astrophysics,
Peking University, Beijing 100871, China}
\affiliation{Department of Physics, University of Guelph, Guelph, Ontario, N1G 2W1, Canada}

\author{Michael LaHaye}
\affiliation{Department of Physics, University of Guelph, Guelph, Ontario, N1G 2W1, Canada}
\author{Huan Yang}
\email{hyang@perimeterinstitute.ca}
\affiliation{Perimeter Institute for Theoretical Physics, Waterloo, Ontario N2L 2Y5, Canada}
\affiliation{Department of Physics, University of Guelph, Guelph, Ontario, N1G 2W1, Canada}
\author{B\'{e}atrice Bonga}
\affiliation{Institute for Mathematics, Astrophysics and Particle Physics,Radboud University, 6525 AJ Nijmegen, The Netherlands}

\begin{abstract}
Spin-induced quadrupole moments provide an important characterization of compact objects, such as black holes, neutron stars and black hole mimickers inspired by additional fields and/or modified theories of gravity.  Black holes in general relativity have a specific spin-induced quadrupole moment, with other objects potentially having differing values.  Different values of this quadrupole moment lead to modifications of the spin precession dynamics, and consequently modifications to the inspiral waveform.  Based on the spin-dynamics and the associated precessing waveform developed in our previous work, we assess the prospects of measuring spin-induced moments in various black hole, neutron star, and BH mimicker binaries. We focus on binaries in which at least one of the objects is in the mass gap (similar to the $2.6 M_\odot$ object found in GW190814). We find that for generic precessing binaries, the effect of the spin-induced quadrupole moments on the precession is sensitive to the nature of the mass-gap object, i.e., whether it is  a light black hole or a massive neutron star. So that this is a good probe of the nature of these objects. For precessing BH mimicker binaries, this waveform also provides significantly tighter constraints on their spin-induced quadrupole moments than the previous results obtained without incorporating the precession effects of spin-induced quadrupole moments. We apply the waveform to sample events in GWTC catalogs to obtain better constraints on the spin-induced quadrupole moments, and discuss the measurement prospects for events in the O$4$ run of the LIGO-Virgo-KAGRA Collaboration.
\end{abstract}

\maketitle

\section{Introduction} 
In the past seven years, the LIGO-Virgo-KAGRA Collaboration (LVK) has detected more than one hundred binary black hole (BBH) merger events, and a handful of events involving neutron stars (NS) (black hole-neutron star (BH-NS) or binary neutron star systems) \cite{GWTC-1,GWTC-2,GWTC-2.1,GWTC-3}.  In the event catalogs, if the gravitational wave (GW) measurement for the mass of an object within the binary is greater than $5 M_\odot$, the object has been identified as a ``BH" by convention.  Similarly, if the mass is less than $2 M_\odot$ (less than $3 M_\odot$ in GWTC-3 \cite{GWTC-3}), it is identified as a NS.  \\

While this classification system is convenient for bookkeeping purposes, it comes with two inherent issues.  First, if the mass distributions of BHs and NSs overlap, we potentially misidentify objects if we only use their masses.  Second, this system fails to say anything about objects lying between these bounds, in the so-called mass gap.  With the unexpected discovery of the $2.6 M_\odot$ object in GW190814 (which can be either a heavy NS \cite{GW190814_BH1,GW190814_BH2} or a light BH \cite{GW190814_NS1,GW190814_NS2,GW190814_NS3,Yang:2017gfb}), we are forced to confront this second issue if we want to determine the nature of this and similar objects.  The nature of these objects can provide insight into their formation mechanism. For example, these objects may also appear in extreme mass ratio inspirals as relevant sources for space-borne gravitational wave detection \cite{Pan:2021lyw}. Their relative abundance in ``wet" (accretion-disk-assisted) \cite{Pan:2021ksp,Pan:2021oob} and ``dry" (scattering-assisted) \cite{Babak:2017tow,Gair:2017ynp} formation channels can be used to constrain supernovae explosion mechanisms,  which is related to possible delayed fall-back accretion that strongly affects the remnant mass. Being able to classify the nature of mass-gap objects correctly is increasingly important.\\

In principle, there are different methods to distinguish between a mass gap NS and BH. While a massive NS has an electromagnetic (EM) counterpart (such as short gamma-ray emission and/or kilonova emission), a BH does not. Consequently, we may distinguish between the two on the basis of the signature of an EM counterpart \cite{Abbott_2017, Graham_2023}. This is weighted by the fact that EM counterpart detection is not always available (i.e., see the EM followups for GW190814 \cite{EM_follow-up1,EM_follow-up2,EM_follow-up3,EM_follow-up4,EM_follow-up5}) either due to faint emission sources or poor sky localization capacities.  The ability to probe the nature of mass-gap objects may also be compromised as the EM signature seems to be greatly influenced by the eccentricity, spin and mass ratio of the system. These are often not accurately constrained by the gravitational wave measurement, in part because a large portion of the parameter space is less explored from the modeling perspective. \\

Hence, we need a method of distinguishing between a massive NS and a light BH via the gravitational waveform alone.  There are several potential gravitational-wave observables that can distinguish between these objects: the tidal deformability, the horizon absorption (HA), and the spin-induced quadrupole moment (SIQM).\\

The most promising observable for lower-mass ($\le 2 M_\odot$) objects should be the (dimensionless) tidal Love number, which is constrained to be $\Lambda(1.4M_\odot)\le 800$ for the low-spin prior in GW170817  \cite{GW170817,GW170817_eos}. However, it is known that the tidal Love number drops dramatically with increasing masses - for objects with masses reaching up to 2.6 $M_\odot$, the dimensionless tidal Love number is $\mathcal{O}(1)$ with an uncertainty marginally achievable by third generation detectors \cite{Tanja_2018,Tanja_2023,Yagi_2013,Castro_2022}.
The HA effect (sometimes referred to as tidal heating), on the other hand, generally enters the waveform phase at $2.5$ post-Newtonian (PN) order for spinning BHs and $4$PN for nonspinning BHs. According to the discussion in \cite{Saketh_2022_horizon} (see also Sec.\ref{subsec:pe} in this work), the corresponding effect is smaller than those of other observables.
As a result, it appears that measuring the SIQM is the only viable approach to discern the nature of mass-gap objects. Due to the quadrupole-curvature coupling, the SIQM affects both the orbital dynamics in the source frame from the aligned-spin induced effect (AI), and the time-dependent precession of the orbital plane related to spin precessions, known as the precessing-induced (PI) effect. (See Sec.\ref{subsec:wf-model} for a detailed discussion.) \\

In this work we show that the PI effect, at least for some configurations, provides a more sensitive probe on the SIQM than the AI effect discussed in previous works \cite{Possion_1998,Arun:2017,Arun:2008kb, Abbott_2017,Narikawa_2021}. In our previous work, we have developed a general precession module, which semi-analytically describes  the generic PI effect for the first time\cite{LaHaye_2022}. Here, we apply our waveform model to investigate the measurement uncertainties of the SIQM coefficient $\kappa$ for various kinds of compact binaries, using Markov Chain Monte Carlo (MCMC) simulations. The detectability prospects are promising, assuming the sensitivity of the A$^\#$ detector. \\



The SIQM of an extended object can be written generally as $Q = -\kappa\, \chi^2 m^3$ \cite{Possion_1998}, where $Q$ is the quadrupole moment scalar, $m$ is the object's mass, $\chi$ its dimensionless spin, and $\kappa$ a dimensionless constant that is one for BHs, and typically greater than one for NSs or other objects \cite{Tanja_2018,Herdeiro_2014,Baumann_2019,Chia_2020}. As the name, spin-induced moment, suggests, the quadrupole moment is a result of the deformation to the object induced by spin.  This means that the most crucial factor affecting the measurement accuracy of $\kappa$ is the magnitude of the spins of the compact objects -- if the spins are zero (or small) the waveform is completely (largely) insensitive to the value of $\kappa$. To make it clear, we use $\kappa_i$ for the $i$th object, while $\kappa$ refers to the generic SIQM.\\

Fortunately, we have reason to suspect that objects in the mass gap have relatively large spin, making them prime candidates for measuring the SIQM.  While there is no definitive observational evidence in favor of a large spin in GW$190814$, current constraints on the spin of the mass-gap object found in GW$190814$ are weak due to limited signal-to-noise ratio (SNR) \cite{GW190814}, so a large spin is not prohibited by this event either. In general relativity, a spinning BH should have dimensionless spin $\chi:= |\Vec{S}|/m^2 \le 1$ to prevent the formation of a naked singularity. For extended objects this is not an issue and consequently $\chi>1$ is allowed \cite{Herdeiro_2014}. For mass-gap objects of the ``NS"-type, a high-spin would help to explain why its mass can exceed the Tolman-Oppenheimer-Volkoff limit of nonrotating stars without collapsing to a BH.  Such large spins are supported by various theoretical models as well. One example of this is in differentially rotating NSs, although these are unstable \cite{Duez_2004,Giacomazzo_2011,Iosif_2021}.  More realistically, numerically solving Einstein's equations of uniformly rapidly rotating NSs shows that $\chi$ could reach up to 0.7.  Another example of objects with $\chi>1$ is quark stars, which could have a spin significantly greater than 0.7 \cite{KaWai_2011}.  For a ``BH"-type mass-gap object, its formation is likely associated with accretion (i.e., from supernovae with significant fall-back accretion \cite{belczynski2012missing} and in active galactic nuclei \cite{Pan:2021xhv}) and merger processes (i.e., being the merger product of a binary neutron star coalescence \cite{Sarin:2020gxb,Kastaun_2013}). Therefore, it is reasonable to include the high-spin possibility of mass-gap objects, so we have used a flat prior for the spin magnitude (from $0$ to $1$) in our simulations.\\

In addition to probing the nature of mass-gap objects, measuring the SIQM also provides key information on testing the existence of BH mimickers, such as boson stars \cite{boson_1997,boson_2017}, gravastars \cite{gravastar_2004,gravastar_2015,gravastar_2023,Yang:2022gic}, anti de-Sitter (AdS) bubbles\cite{bubbles_2017,Danielsson:2021ykm} and alternative BH solutions in modified theories of gravity \cite{Herdeiro_2014}. In the work of testing general relativity using GW events by the LVC \cite{tgr_O1,tgr_O2,GWTC3-tgr}, the spin-induced moments have been constrained for individual events and with combined posteriors, using waveform models with only the non-precessing effect of SIQMs (AI effect only). Here, we re-perform the analysis incorporating the PI effect, and indeed find better constraints on $\kappa$ for sample gravitational wave events in GWTC catalogs.
We also investigate the measurement prospects for the events in the LVK O4 science run. \\

This paper is structured as follows. In Sec.~\ref{sec:mass-gap}, we first develop a waveform model by including both the precessing and non-precessing SIQM effect and the HA effect based on existing IMRPhenomXPHM \cite{Pratten_2020imr1,Pratten_2020imr3,Garcia-Quiros_2020imr2}. Then we consider four mass-gap binaries with different mass ratios and spins configurations and calculate the cumulative distribution function (CDF) of the {\it Mismatch} in Sec.~\ref{subsec:binary}. Finally, we configure four injected waveforms with a medium value of the {\it Mismatch} from the PI effect alone for each case and run MCMC simulations. The results are shown in Sec.~\ref{subsec:result}. In Sec.~\ref{sec:mimickers}, we demonstrate the detectability of BH mimickers from GW events with O4 sensitivity.  \footnote{ The codes implemented in this work, as well as in our previous one \cite{LaHaye_2022}, are publicly available on GitHub: https://github.com/GWLyu/SIQM.} \\

\section{Mass-gap Binaries}\label{sec:mass-gap}
The primary target sources of this work are compact binaries including mass-gap objects, and binaries containing BH mimickers. With the inclusion of the PI effect, we shall show that the SIQM $\kappa$ can be better measured/constrained, which in turn can be used to infer the nature of the mass-gap objects and/or test the existence of BH mimickers.

\subsection{Waveform Model}\label{subsec:wf-model}
The SIQM effect enters the GW waveform model in two ways: through a non-precessing and a precessing part. For the non-precessing part of the waveform (AI effect), it is known that the SIQM introduces phase corrections starting at 2PN order, and further corrections have been worked out up to 3.5PN order \cite{Possion_1998,Arun:2017,Arun:2008kb}\,,
\begin{align}
\delta \phi_{\rm QM} =& \frac{3}{128 \,\eta\, v^5} \left (\lambda_{\mathrm{2PN}}\, v^4 + \lambda_{\mathrm{3PN}}\, v^6\right )\,,\\
\lambda_{\mathrm{2PN}} =& -10\,\sigma_{\mathrm{QM}} = 25 \sum^2_{i=1}\left [3 (\hat{\chi}_i \cdot \hat{L})^2-1 \right ] \frac{Q_i}{m_i M^2} \,,\\
\lambda_{\mathrm{3PN}} =& -\frac{5}{84}\sum^2_{i=1}(9470+8218 X_i-2016 X_i^2)\frac{Q_i}{m_i M^2} \,,
\end{align}
where $M=m_1+m_2$, $X_i=m_i/M$ (this is referred to as $\mu_i$ in our previous work), $\eta=X_1 X_2 = m_1 m_2/M^2$ is the symmetric mass ratio. The PN expansion parameter is $v=(M\omega_{\rm orbit})^{1/3}=(\pi M f)^{1/3}$ with $\omega_{\rm orbit}=\pi f$ being the orbital frequency of the binary system, and $f$ is the GW frequency. $Q_i$ is SIQM of the $i$th object. $\hat{\chi}_i$ is the unit vector along the spin direction of the $i$th object and $\hat{L}$ is the unit vector along the direction of the orbital angular momentum.\\ 

Such phase modulations due to SIQM have been incorporated in IMRPhenomPv2 to obtain constraints on $\kappa$ for BH mimickers in the GWTC catalogs \cite{tgr_O2,GWTC3-tgr}. In this work, we  adopt the state-of-art IMRPhenomXPHM model. The differences between these two models are as follows. First, the IMRPhenomPv2 waveform is based on the next-to-next-to-leading order single-spin PN approximation \cite{NNLO}, while IMRPhenomXPHM is based on the double-spin MSA approach \cite{MSA} in order to map the aligned-spin waveform modes in the co-precessing $L$-frame to the inertial $J$-frame. Second, the IMRPhenomPv2 waveform uses a non-spinning 2PN approximation to the orbital angular momentum, while IMRPhenomXPHM uses an aligned-spin 4PN approximation including spin-orbit contributions. The MSA approach performs much better as an approximation to the Euler angles as shown in Fig. 3 of \cite{Pratten_2020imr3}.   Finally, the IMRPhenomPv2 waveform does not include higher-order multipoles in the radiation.\\

For precessing binaries, the SIQM affects the precession of the orbital angular moment and spins according to the precession frequency \cite{Klein_2021} up to 2PN order.
\begin{align}
    \Vec{\Omega_i} =\frac{1}{X_i\, v^5} \left[\left(\frac{1}{2}X_i + \frac{3}{2}-\frac{3}{2}v\hat{L}\cdot(\kappa_i\,\Vec{s_i}+\Vec{s_j})\right)\Hat{L} + \frac{1}{2}v\,\Vec{s_j}\right] \,,
\end{align}
The equation of precession for spin $\Vec{s_i}$ is $\dot{\Vec{s_i}} = \Vec{\Omega}_i\times\Vec{s_i}$, where $\Vec{s_i}= \Vec{S_i}/X_i$ are the reduced spin parameters.

As a result, an extended object with $\kappa \neq 1$ generally leads to precession dynamics different from a BH, even if it has the same mass and spin as the BH. This precession effect of the SIQM was first incorporated into a precessing waveform model in \cite{LaHaye_2022}, where we developed novel procedures to efficiently solve the spin dynamics on the gravitational radiation reaction timescale. This is nontrivial as one of the conserved quantities for the $\kappa=1$ spin dynamics is no longer conserved for generic $\kappa$, making it difficult to obtain the algebraic solutions of the spin evolution equations. The resulting frequency-dependent rotation of the orbital frame can be applied to a non-precessing waveform to convert it to a precessing waveform model. Currently the waveform includes terms in the precessing dynamics up to 2PN order, but higher-order PN terms are currently available and can be included in our method easily. In this work, we apply the framework in \cite{LaHaye_2022} and modify the precession module of the phenomenological inspiral-merger-ringdown waveform model IMRPhenomXPHM \cite{Pratten_2020imr1,Pratten_2020imr3,Garcia-Quiros_2020imr2} so that it applies for generic $\kappa$ with both AI and PI effects.  This waveform model also includes multiple GW harmonics beyond the $22$ modes which are sensitive to the rotation of the orbital plane. Hence, we include $(l,m)=(2,2), (2,1),(3,3)$ modes in this work and ignore other modes to save the computational resources since the ignored ones have limited contribution to the mismatch. In addition, a SpinTaylor precession version of IMRPhenomXPHM also exists and is publicly available in LALSuite \cite{lalsuite}. It accounts for the SIQM in the twisting-up procedure by performing a SpinTaylor PN evolution to generate the precession angles \cite{colleoni2023imrphenomxpnrtidalv2}. \\

Besides the SIQM effects, we also allow the HA effect to be turned on or off depending on the nature of the compact object, i.e., a BH or a NS \cite{Tagoshi_1997_horizon,Alvi_2001_horrizon,Saketh_2022_horizon}. For spinning BHs the HA effect enters the waveform at 2.5PN order and for non-spinning BHs it enters at the 4PN order. For simplicity, we have only included the 2.5PN and 3.5PN order contribution, as given by \cite{Mukherjee_2022,Datta_2021}
\begin{align}
\delta \phi_{\rm HA} =& \frac{3}{128 \,\eta\, v^5} (\lambda_{\mathrm{2.5PN}}\, v^5 +\lambda_{\mathrm{3.5PN}}\, v^7)\,,\\
\lambda_{\mathrm{2.5PN}} =& -\frac{10}{9}H_{\rm eff}\,(3\log(v)+1)\,,\\
\lambda_{\mathrm{3.5PN}} =& -\frac{5}{168}(952\eta+955)H_{\rm eff}\,,\\
H_{\rm eff} =& \sum_{i=1}^2 X_i^3(\hat{L}\cdot\hat{\chi}_i)\,\chi_i\,(3\chi_i^2+1)\,,
\end{align}
where we have assumed that the energy flux is fully absorbed if the horizon exists and no absorption if there is no horizon.\\

We have not incorporated the tidal deformability of the mass-gap object into the waveform model, since it is expected to be small. For example, the tidal Love number of a $2.6 M_\odot$ object is expected to be $\mathcal{O}(10)$ \cite{Tanja_2018}, which is only marginally detectable by third-generation gravitational wave detectors \cite{Tanja_2018,Tanja_2023,Yagi_2013,Castro_2022}. In other words, we are discussing the problem of how to probe the nature of a compact object when the tidal deformability measurement is not informative for this task. Thus, the tidal deformability has limited influence on the constraint of SIQM for mass gap objects, but could potentially be important for BH mimickers, where the tidal deformability can be larger.

\subsection{Binary Configurations}\label{subsec:binary}
To qualitatively address the detectability of the SIQM coefficient $\kappa$ for mass-gap binaries, we consider four sample cases with different mass ratios and spin magnitudes, as they are the key factors determining the measurement accuracy of $\kappa$. The detailed set of parameters is summarized in Table~\ref{tab1}. The ``C2"-type binary, consisting of a BH and a mass-gap object, has component masses and primary spin consistent with the measured parameters of GW190814. The injection value of $\kappa$ for the more massive object is one, since it is a BH. The injected value of the mass-gap object is two, which is consistent with the number discussed in \cite{kappa2}. The ``C1"-type binary is similar to the ``C2''-type except that this primary spin is assumed to be significant. The ``C3" binary represents a so-far undetected type of binary containing double mass-gap objects. The ``C4" binary represents another undetected possibility with a NS and a mass-gap object. \\

For each type of binary, we would like to compare the detectability of the HA, the non-precessing and precessing SIQM effects, as a way to determine which effect provides the most sensitive probe on the nature of the mass-gap object. For this comparison, we compute the waveform {\it Mismatch}, which characterizes the difference between two waveforms. Mathematically the {\it Mismatch} is defined as $1-match$, where $match$ is the inner product of two normalized waveforms and maximized over coalescence time $t_c$ and reference phase $\phi_c$ \cite{Babak_2013}\,,
\begin{equation}\label{eq:mimatch}
\mathcal{M}(h_1, h_2) = 1 - \max_{t_c,\phi_c}\, \frac{<h_1, h_2>}{\sqrt{<h_1, h_1>}\,\sqrt{<h_2, h_2>}}\,,
\end{equation}
where $<.\;,\,.>$ is the noise-weighted inner product defined as \cite{Finn_1992}
\begin{equation}
  \left<a(t)|b(t)\right> = 2\int_{f_{\rm low}}^{f_{\rm high}} \frac{\tilde{a}^*(f)\tilde{b}(f) + \tilde{a}(f)\tilde{b}^*(f)}{S_n(f)} df\,,
\end{equation}
Here $*$ refers to a complex conjugate, $S_n(f)$ is the one-sided detector-noise power spectral density (PSD) of given detectors, $f_{\rm low}$ and $f_{\rm high}$ are the low and high frequency cutoff, respectively.
The {\it Mismatch} is computed for all cases C1 to C4, between a waveform assuming the mass-gap object is a BH ($\kappa=1$) and a waveform with one of the three effects implemented to the mass-gap object.
\\

\begin{figure*}[ht]
	\centering
    \hspace{-4pt}\includegraphics[width=0.49\textwidth]{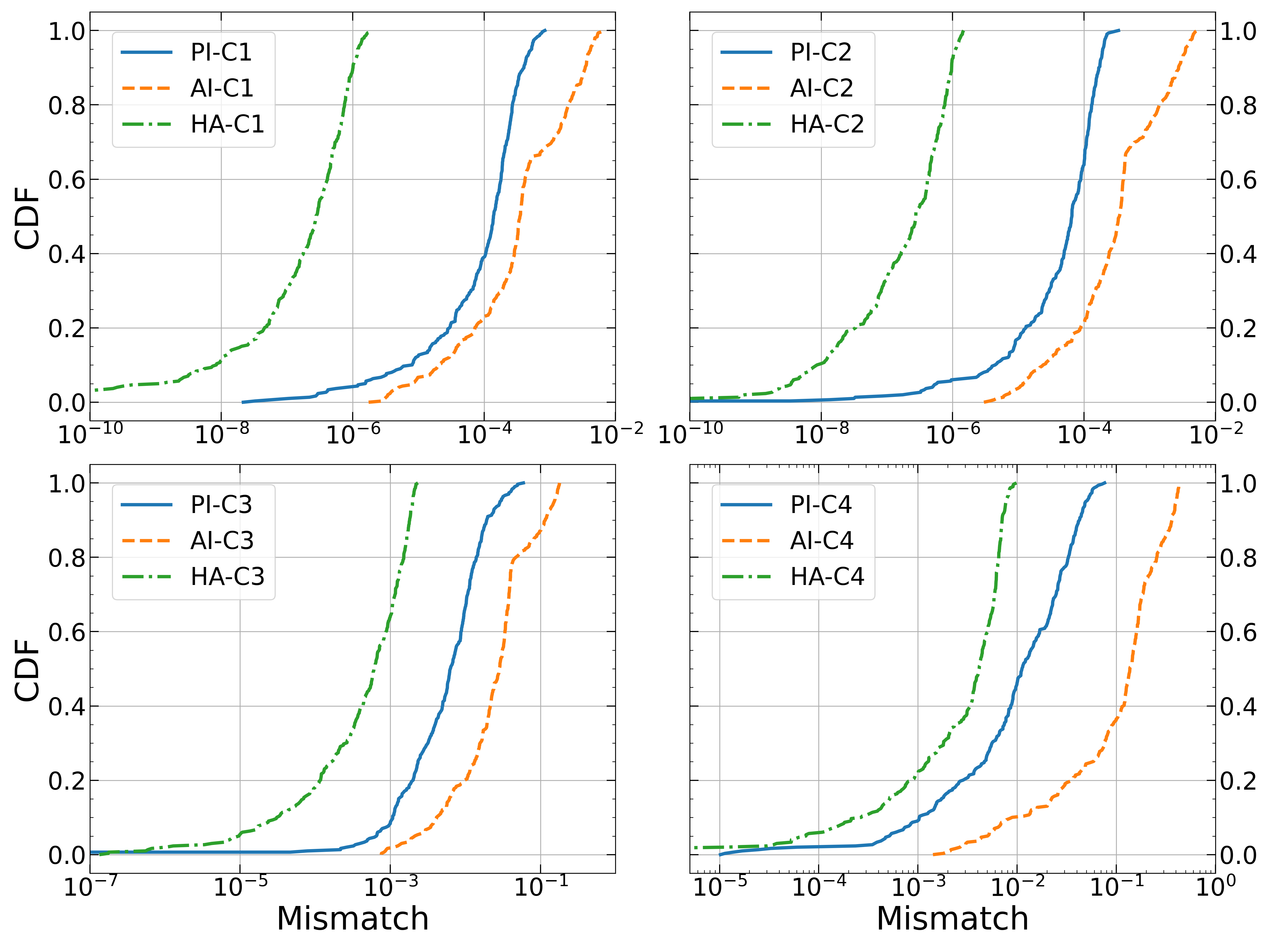}
    \hspace{8pt}\includegraphics[width=0.49\textwidth]{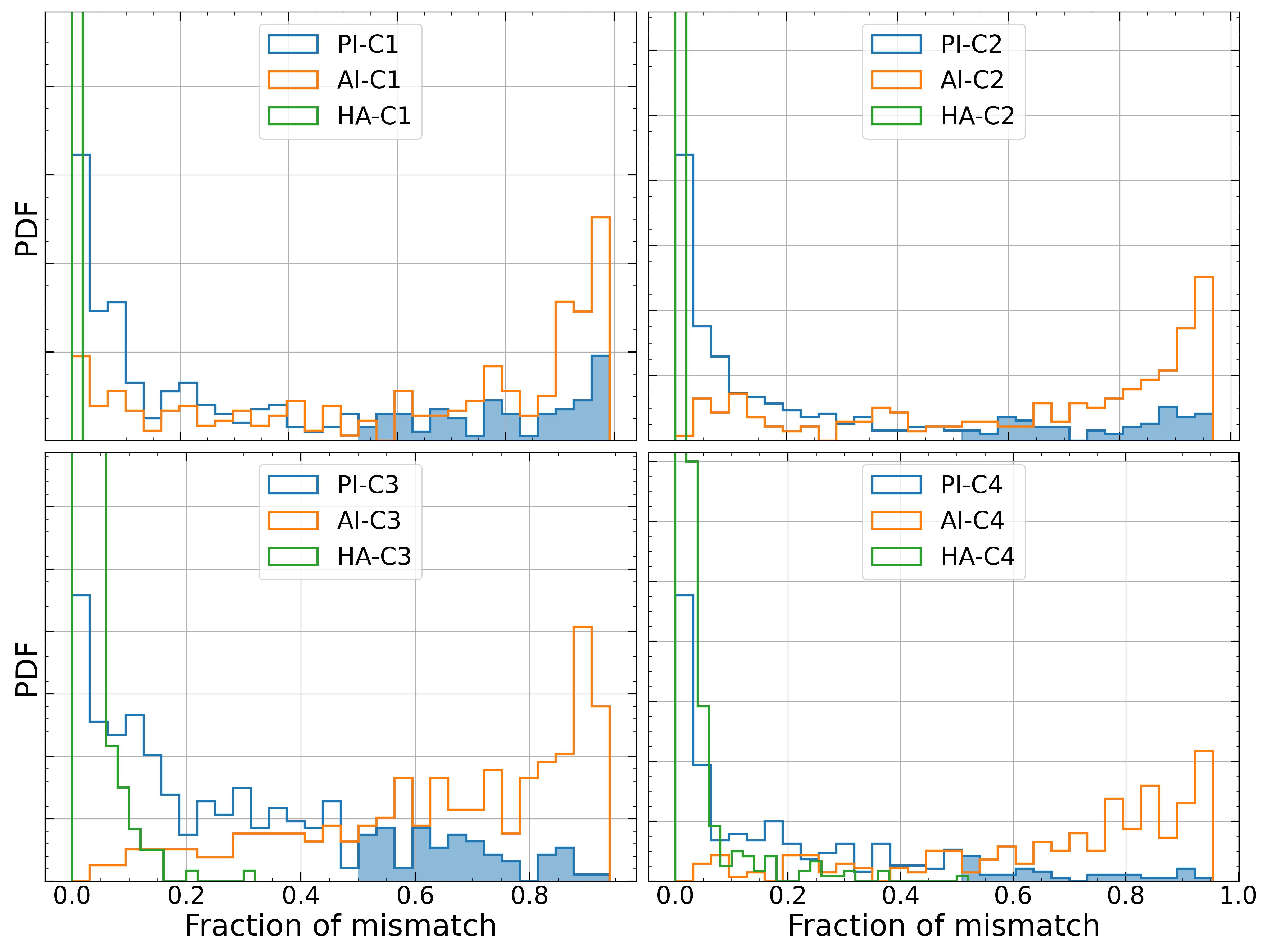}
	\caption{Left: the CDF of each effect for four different cases. PI is an abbreviation for precession induced, AI refers to aligned-spin induced while HA is horizon absorption effect. The $Mismatch$es of C1 and C2 are two order of magnitudes smaller than C3 and C4 because their mass ratios ($q=m_1/m_2>1$) are much greater than C3 and C4. Right: The distribution of fraction of mismatch ($F_i$) defined in En.~\ref{eq:fraction} of each effect for four cases. The histogram is normalized for individual effects. We could see the shaded areas which demonstrate the fraction of samples with more than $50\%$ mismatch contribution from the PI effect are about $31\%\,, 23\%\,, 21\%$\,, and $14\%$\,, for cases C1 to C4, respectively.}
	\label{fig:cdf_frac}
\end{figure*}

The initial spin orientations determine the degree to which different effects induce differences in the waveform. As an example, for nearly aligned spin systems the PI effect is minimal (because the precession in general is minimal). In order to make a fair comparison, we have generated random initial spin orientations according to uniform distributions in the sky.
Fig.~\ref{fig:cdf_frac}\footnote{Here we do not consider the detector response, which means we make use of GW polarization $h_{+,\times}$ directly and weighted with A$^\#$ PSD.} shows the CDF of the {\it Mismatch} according to these randomly generated polar angles and azimuthal angles of spins over the unit sphere, and the inclination angles from a cosine distribution. \\

\renewcommand{\arraystretch}{1.5}
\begin{table}[t]
\centering
\begin{tabularx}{0.45\textwidth} {>{\centering\arraybackslash}X >{\centering\arraybackslash}X | >{\centering\arraybackslash}X | >{\centering\arraybackslash}X | >{\centering\arraybackslash}X | >{\centering\arraybackslash}X}
\hline
\hline
 & &  C1 & C2 & C3 & C4 \\
\hline
\multicolumn{2}{c|}{$m_1(M_\odot)$} & 23 & 23 & 3.0 & 2.6 \\

\multicolumn{2}{c|}{$m_2(M_\odot)$} & 2.6 & 2.6 & 2.6 & 1.3 \\

\multicolumn{2}{c|}{$\chi_1$} & 0.6 & 0.07 & 0.7 & 0.7 \\

\multicolumn{2}{c|}{$\chi_2$} & 0.7 & 0.7 & 0.7 & 0.05 \\
\hline
\multicolumn{2}{c|}{$\kappa_1$} & 1 & 1 & 1 & 2 \\
\hline
\multicolumn{2}{c|}{$\kappa_2$} & 2 & 2 & 2 & 6 \\
\hline
\hline
\end{tabularx}
\caption{Configurations of four cases. Each case has a $2.6M_\odot$ mass-gap object with $\kappa=2$, and $\kappa=1$ for BHs (i.e., all objects above with masses larger or equal to $3M_\odot$).}
\label{tab1}
\end{table}

The left panel of Fig.~\ref{fig:cdf_frac} shows that, for all four cases, the HA effect produces the least mismatch and the AI effect leads to the highest overall mismatch. The right panel, on the other hand, compares the fraction of mismatch as defined by
\begin{align}\label{eq:fraction}
F_i : =\frac{\mathcal{M}_i}{\mathcal{M}_{\rm PI}+\mathcal{M}_{\rm AI}+\mathcal{M}_{\rm HA}}, \quad i={\rm PI, AI, HA}
\end{align}
The PI effect is the dominant mechanism for producing mismatch for approximately $20\%$ of the spin configurations, where spin precession is significant.\\

The mismatch due to HA is consistent with the analysis in \cite{HA01,HA02}, where the phase modulation due to the HA effect is shown to be extremely small.  Comparisons can also be made between the mismatch of different cases. For example, for the mismatch introduced by PI effects, the values for C1 binaries are generally larger than those  of C2 binaries, because of the larger spin assumed for object one in C1. The mismatch also seems to be higher for comparable mass-ratio systems, as the magnitude of mismatch of C3 binaries is significantly higher than C1 binaries. In the first three cases, this is due to the fact that the effect of the SIQM on the waveform is suppressed by the mass ratio, since only the smaller object has $\kappa\neq 1$.  In the fourth case, since both objects have $\kappa\neq 1$, the effect is large, regardless of mass ratio since the large object's effect is not suppressed by the mass ratio.  If the binary contains a normal NS, such as the C4 binaries, the associated $\kappa$ may be significantly greater than that of the mass-gap objects \cite{Tanja_2018}. Although the underlying spin of the NS is assumed to be small $\chi_2=0.05$, the resulting mismatch is still mostly in the range between $10^{-2}$ to $10^{-1}$.\\

While the mismatch is informative for comparing effects very roughly, the {\it Mismatch} study alone is not able to provide quantitative measures on the measurement accuracy of $\kappa$ across the different scenarios. In the next section we use the Bayesian Inference method, together with MCMC parameter estimation to obtain posterior distribution of system parameters, including $\kappa$, for selected sets of scenarios. This task can not be done simultaneously for {\it all} spin configurations shown in Fig.~\ref{fig:cdf_frac} because of the large computational cost associated with these MCMC simulations. In order to pick the most representative cases, we first adopt the median mismatch configurations from the CDF plots  as injected system parameters to perform the Bayesian inference.\\

As we will show in Sec.~\ref{subsubsec:PIAI_effects}, in the simulations having the same level of mismatch does not necessarily imply the same level of measurement uncertainty of $\kappa$. Although the mismatch contribution from AI effect is significantly greater than PI effect for all four types of systems, the MCMC simulations actually show that the posterior distribution of $\kappa$ is more sensitively determined by the PI effect rather than the AI effect, even if their corresponding mismatch is comparable. For almost all cases where the spin precession is significant, the contribution to the posterior distribution of $\kappa$ from AI effect is smaller than the PI effect. As a result, later on when we pick sample spin configurations to estimate the measurement uncertainty of $\kappa$,  we select the median mismatch configurations from the PI effect alone.

\subsection{Parameter Estimation}\label{subsec:pe}
In order to compute the measurement uncertainty, we shall apply the Bayesian inference method \cite{Bayes2019,Bayes2021}, which is developed based on Bayes' theorem and MCMC. According to the Bayes' theorem, given GW data $d$ and hypothesis $\mathcal{H}$, the posterior distribution is given by
\begin{equation}
p(\mathcal{\boldsymbol{\vartheta}} | d, \mathcal{H} ) =  \frac{p(d | \boldsymbol{\vartheta}, \mathcal{H})\; p(\boldsymbol{\vartheta} | \mathcal{H})}{p(d | \mathcal{H})} = \frac{p(d | \boldsymbol{\vartheta}, \mathcal{H})\; p(\boldsymbol{\vartheta} | \mathcal{H})}{\int d\boldsymbol{\vartheta}\; p(d | \boldsymbol{\vartheta}, \mathcal{H})\;p(\boldsymbol{\vartheta} | \mathcal{H})}\,,
\end{equation}
Here $p(d | \boldsymbol{\vartheta}, \mathcal{H})$ is the likelihood function while $p(\boldsymbol{\vartheta} | \mathcal{H})$ is the prior on $\mathcal{\boldsymbol{\vartheta}}$. Assuming a stationary Gaussian noise, the log likelihood function $\log p(d|\boldsymbol{\vartheta}, \mathcal{H})$ can be expressed as
\begin{equation}
  \log p(d|\boldsymbol{\vartheta}, \mathcal{H}) =  \log\bar \alpha -\frac{1}{2} \sum_{k}\left< d_k - h_k(\boldsymbol{\vartheta}) | d_k - h_k(\boldsymbol{\vartheta}) \right>\,,
\end{equation}
where the index $k$ refers to different detectors and $\log\bar \alpha$ is the normalization factor while $d_k$ and $h_k(\boldsymbol{\vartheta})$ are the data and waveform templates from given detectors.\\

As explained in Sec.~\ref{subsec:binary}, we shall compute the posterior distributions for selected parameter configurations for all four cases. Among the randomized spin configurations and orbital inclinations in Fig.~\ref{fig:cdf_frac}, we choose the ones corresponding to the median value of PI mismatch in each category since the contribution to the measurement accuracy of $\kappa$ from the AI effect is much weaker than the PI effect, even though this would not be clear from their relative mismatches (see Sec.~\ref{subsubsec:PIAI_effects}). The relevant system parameters are shown in Table~\ref{tab2}, with the rest of parameters consistent with Table~\ref{tab1}. Notice that for each setup, there are 16 or 17 prior parameters in total depending on whether $\kappa$ from the non-$2.6M_\odot$ object is included or not. For all injection configurations the luminosity distance is set to be $d_\mathrm{L}=200$ Mpc, sky location is fixed as $(\alpha, \delta)=(1.0,2.0)$ and polarization angle $\psi=3.0$. The injection values of coalescence phase and coalescence time are chosen randomly, since they have almost no influence on the simulation results.\\

We employ LALSuite \cite{lalsuite} and PyCBC \cite{Biwer_2019} packages with specific modifications to generate waveforms and run MCMCs. Regarding to the MCMC configurations, we adopt the $marginalized\_polarization$ likelihood model developed by PyCBC group. This model numerically marginalizes over polarization angle which will reduce one prior parameter. A modified version of $rwalk$ named $rwalk2$ method is chosen to sample the prior space. Other important setups are: $nlive=5000$, $dlogz=0.1$ and burn-in test is $nacl$ and $max\_posterior$. The auto-correlation length is set to be $nacl=5$ by default and increased accordingly if the posterior distributions do not converge well. A low frequency cutoff $f_{\rm low} = 20$ Hz is set for all simulations and we always set a uniform prior on $\kappa$ with a range which doesn't affect the results. In addition, all the injected signals are generated with the same waveform model as the MCMC realization unless stated differently. \\

\renewcommand{\arraystretch}{1.5}
\begin{table}[t]
\centering
\begin{tabularx}{0.45\textwidth} {>{\centering\arraybackslash}X >{\centering\arraybackslash}X | >{\centering\arraybackslash}X | >{\centering\arraybackslash}X | >{\centering\arraybackslash}X | >{\centering\arraybackslash}X}
\hline
\hline
 & &  C1 & C2 & C3 & C4 \\
\hline
\multicolumn{2}{c|}{$\theta_1$} & 1.05 & 2.37 & 1.07 & 1.43 \\

\multicolumn{2}{c|}{$\theta_2$} & 1.13 & 2.63 & 2.25 & 0.35 \\

\multicolumn{2}{c|}{$\phi_1$} & 0.51 & 0.91 & 2.50 & 5.73\\

\multicolumn{2}{c|}{$\phi_2$} & 0.77 & 4.70 & 1.01 & 2.09 \\
\multicolumn{2}{c|}{$\iota$} & 1.43 & 1.15 & 1.49 & 2.52 \\
\hline
\multirow{2}{*}{SNR} & A$^\#$ & 122 & 97 & 27 & 61 \\
\cline{2-6}
 & CE & 1568 & 1401 & 250 & 736 \\
\hline
\hline
\end{tabularx}
\caption{Parameters corresponding to the median value of PI mismatch for MCMCs. $\theta_i$ and $\phi_i$ are the polar and azimuthal angles, respectively. $\iota$ refers to the inclination angle.  The last two rows show the optimal SNRs for different detector networks.}
\label{tab2}
\end{table}

\begin{figure}[t]
	\centering
    \includegraphics[trim={0cm 1cm 1cm 1cm},clip=true, width=0.48\textwidth]{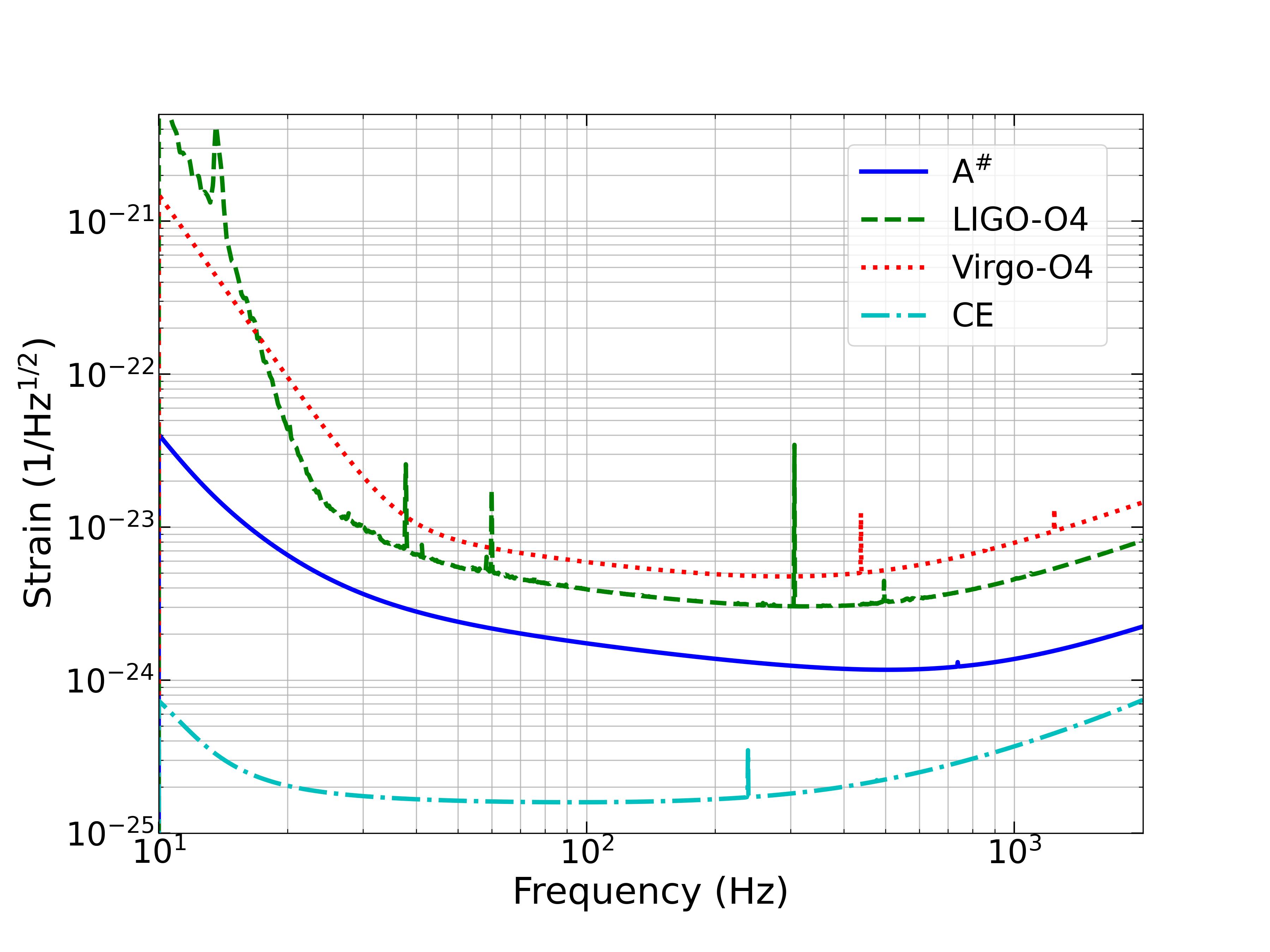}
	\caption{Sensitivity curves of LIGO-O4, Virgo-O4 \cite{LVO4}, A$^\#$ \cite{Asharp}, and CE \cite{CE} detectors.} 
	\label{fig:sensitivity}
\end{figure}

We would like to assess the detectability of the SIQM in different eras of detector development. For this purpose, we have assumed two sample detector networks. We have dubbed the first sample network as the A$^\#$ network consisting of five detectors with A$^\#$ sensitivity located at Hanford, Livingston, LIGO India, Virgo and KAGRA, respectively. Note that  the LIGO detectors have started the O$4$ observing run with expected duration of $18$ months. There will be a following O$5$ observing run scheduled for late 2020s, and a post-O$5$ upgrade -- the A$^\#$ detector using the same LIGO facility. This first sample detector network is intended to show the constraining power of such a detector network. The other sample detector network, dubbed the cosmic explorer (CE) network here, consists of three detectors with CE sensitivity located at Hanford, Livingston, and LIGO India, respectively.  This is chosen to represent the constraining capability of a network of third-generation ground-based detectors; leading concepts for which are the US-led detector Cosmic Explorer \cite{Abbott2017CE,reitze2019cosmic} and European-led detector Einstein Telescope \cite{Punturo:2010zz,hild2008pushing}, of which we have chosen CE for simplicity as a representative third generation detector. The corresponding sensitivity curves for various detectors are shown in Fig.~\ref{fig:sensitivity}.\\

In Table~\ref{tab2}, we display the event SNR for each case, assuming the A$^\#$ network and the CE network respectively. The optimal matched filter SNR  $\rho_{\rm opt}$ is defined as
\begin{equation}
    \rho_{\rm opt} = \sqrt{\sum_k<h_k|h_k>}=\sqrt{4\sum_k\int_{f_{\rm low}}^{f_{\rm high}} \frac{\tilde{h_k}^*(f)\tilde{h_k}(f)}{S_{n,k}(f)} df}\, .
\end{equation}
where the index $k$ refers to the $k$th detector and $h_k$ is the GW strain obtained at the $k$th detector. C1 and C2 cases have higher SNR because they have larger chirp mass for the binaries.  The SNR is also sensitive to the inclination angle of the orbital plane.

\subsection{Simulation Results}\label{subsec:result}
\subsubsection{Significance of different effects}\label{subsubsec:PIAI_effects}
In order to study the impact of the PI and AI effect in constraining $\kappa$, we consider a GW190814-like event (C2) as a sample system (as shown in Table~\ref{tab1} and \ref{tab2}) to perform MCMC simulations. The waveform (modified IMRPhenomXPHM) includes all the effects (PI, AI and HA). Notice that this C2 system already contains significant spin precession, and we choose to use waveforms with the PI or the AI effect for parameter estimation. For completeness, we also include a MCMC study with the TaylorF2 waveform commonly used for aligned-spin systems. For this case, the injected values for the aligned spins $\chi_{1z}$ and $\chi_{2z}$ are assumed to be the same as the magnitude of spins $\chi_1$ and $\chi_2$  in the precessing case.\\

\begin{figure*}[ht]
	\centering
    \hspace{-5pt}\includegraphics[width=0.46\textwidth]{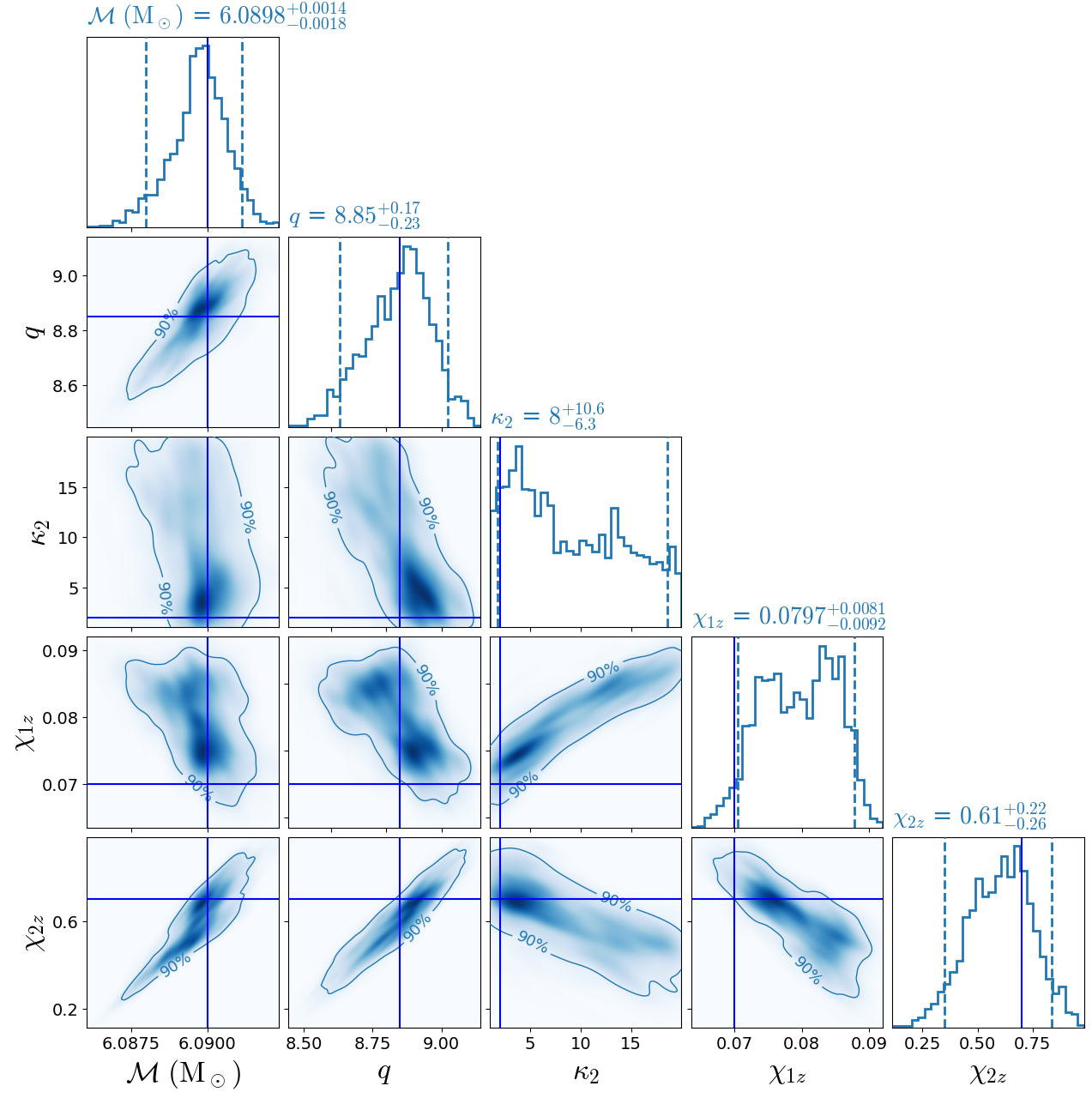}\hspace{10pt}
    \includegraphics[width=0.52\textwidth]{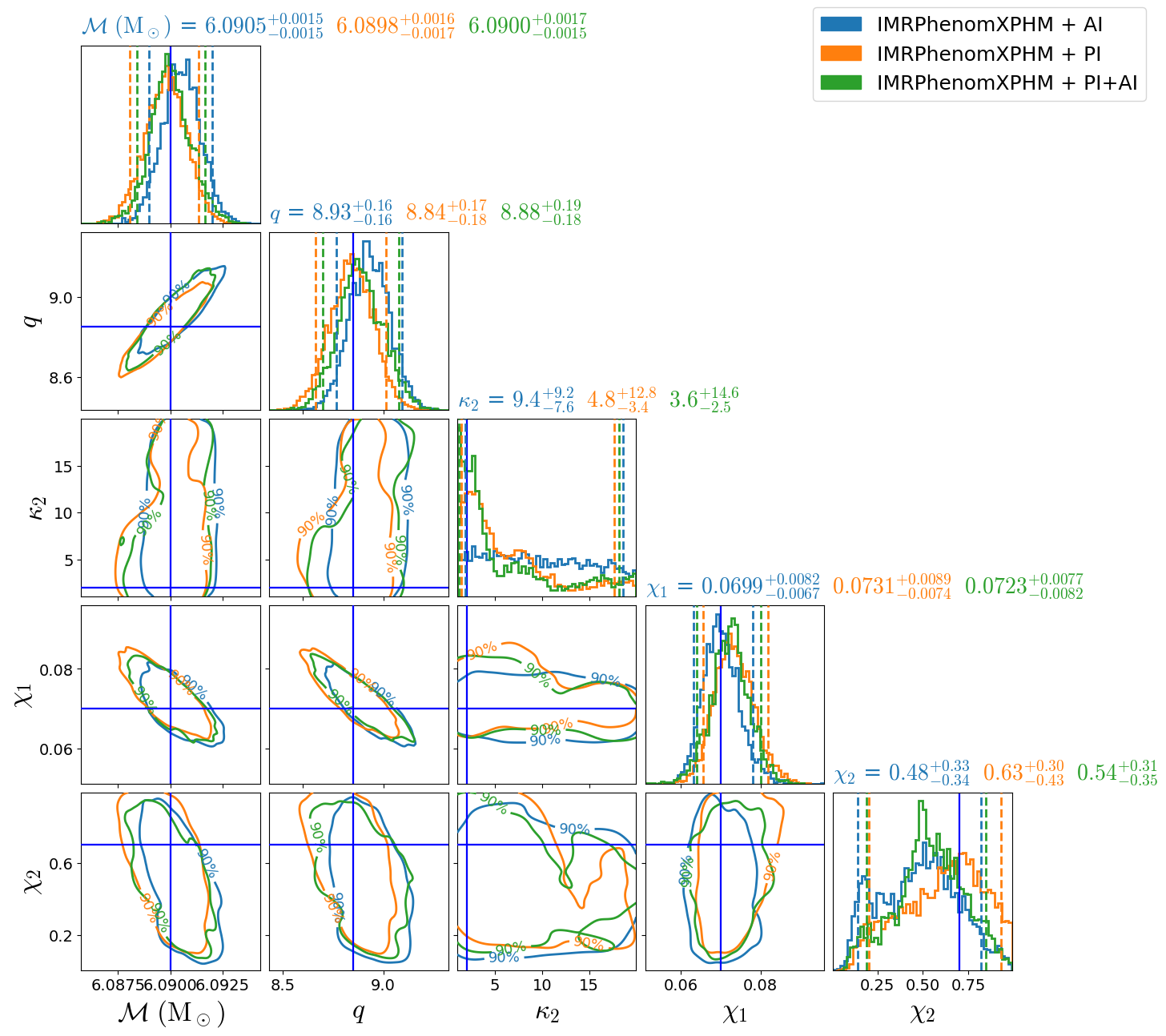}
	\caption{Left: A corner plot shows the degeneracy between SIQM $\kappa_2$ and spins $\chi_{1z}$ and $\chi_{2z}$ with TaylorF2 waveform model. Right: A comparison between AI, PI, or PI+AI effects in precessional waveform models with the same injection while the fractional mismatch of PI is close to AI with $\approx 50\%$ of the total mismatch (Note that there is no PI effect in the TaylorF2 waveform model and we assumed aligned spins in the left plot). The dashed lines in marginalized distributions are the $5\%$ and $95\%$ credible level lines.}
	\label{fig:PIAI_effects}
\end{figure*}

For waveforms including only non-precessing AI effects, it is known that there is a degeneracy between the $\kappa$ and spins $\chi_{1z}$ and $\chi_{2z}$ \cite{Tanja_2018}. We illustrate this point with a simulation (GW190814-like event (like C2) with A$^\#$ network) using the TaylorF2 waveform model. The result  is shown  in the left panel of Fig.~\ref{fig:PIAI_effects} with  a clear degeneracy observed between $\kappa$ and the spin parameters, which generally degrades the measurement accuracy of $\kappa$. This degeneracy is no longer present for precessing systems as shown in the right panel of Fig.~\ref{fig:PIAI_effects}. However, even if there is no degeneracy, it seems that the precessing waveform model including only the AI effect has relatively worse measurement accuracy on $\kappa$. In Fig.~\ref{fig:PIAI_effects}, we also present the posterior distribution of parameters with the PI effect included in the waveform model. We find that the constraint on $\kappa$ becomes much tighter. Notice that the underlying spin configuration is chosen such that the relative contribution to the total mismatch from PI and AI effects is roughly $50\% : 50\%$. The result shows that for the same level of mismatch, PI effects tend to have much better correlation with the measurement uncertainty of $\kappa$. Therefore, when we pick a sample binary parameter to estimate the ``typical" measurement uncertainty of $\kappa$ for different types of systems and for different spin configurations, it should be more appropriate to choose injected waveform parameters (shown in Table~\ref{tab2}) with a median value of mismatch for the PI effect alone (because it is more relevant for the $\kappa$ uncertainty) rather than the total mismatch as shown in Fig.~\ref{fig:cdf_frac}.\\

In addition, for binaries with two non-BH objects, it is noted in \cite{Tanja_2018} that the degeneracy between $\kappa_1$ and $\kappa_2$ severely compromises the possibility of individually measuring $\kappa_{1,2}$. Including higher PN correction terms does not significantly improve the situation. We find that the $\kappa$ for both mass-gap objects (C3) can be individually measured without being affected by degeneracies, thanks to the precessing SIQM effect and the relatively high-spin magnitude ($\sim 0.7$) assumed for the mass-gap objects. \\

\begin{figure*}[ht]
	\centering
    \includegraphics[width=0.48\textwidth]{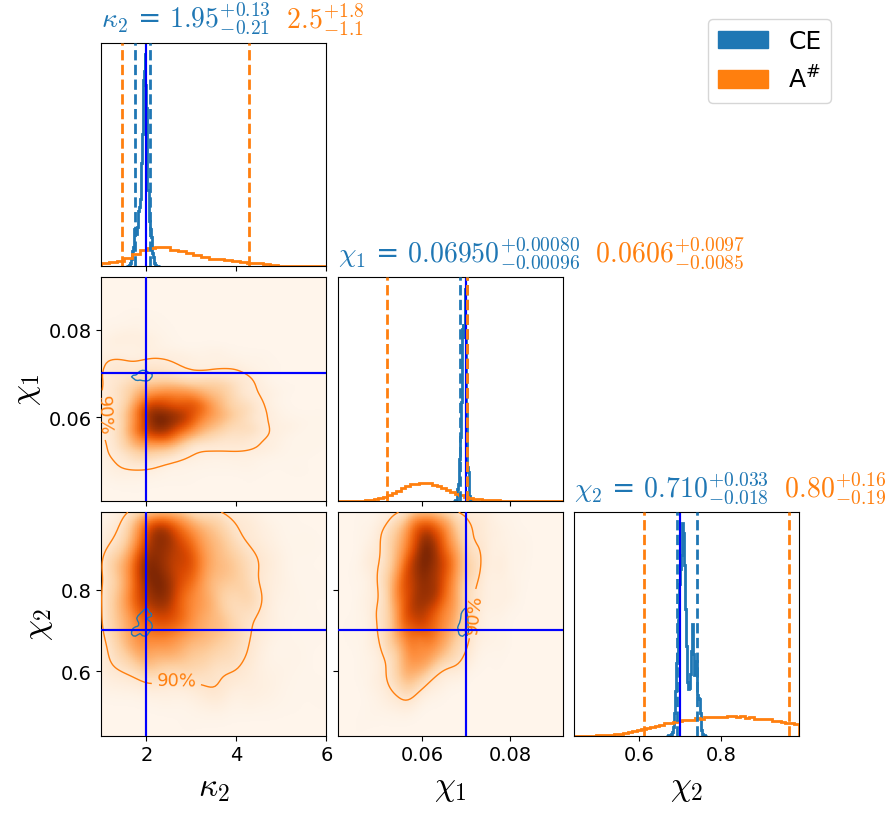}
    \includegraphics[width=0.51\textwidth]{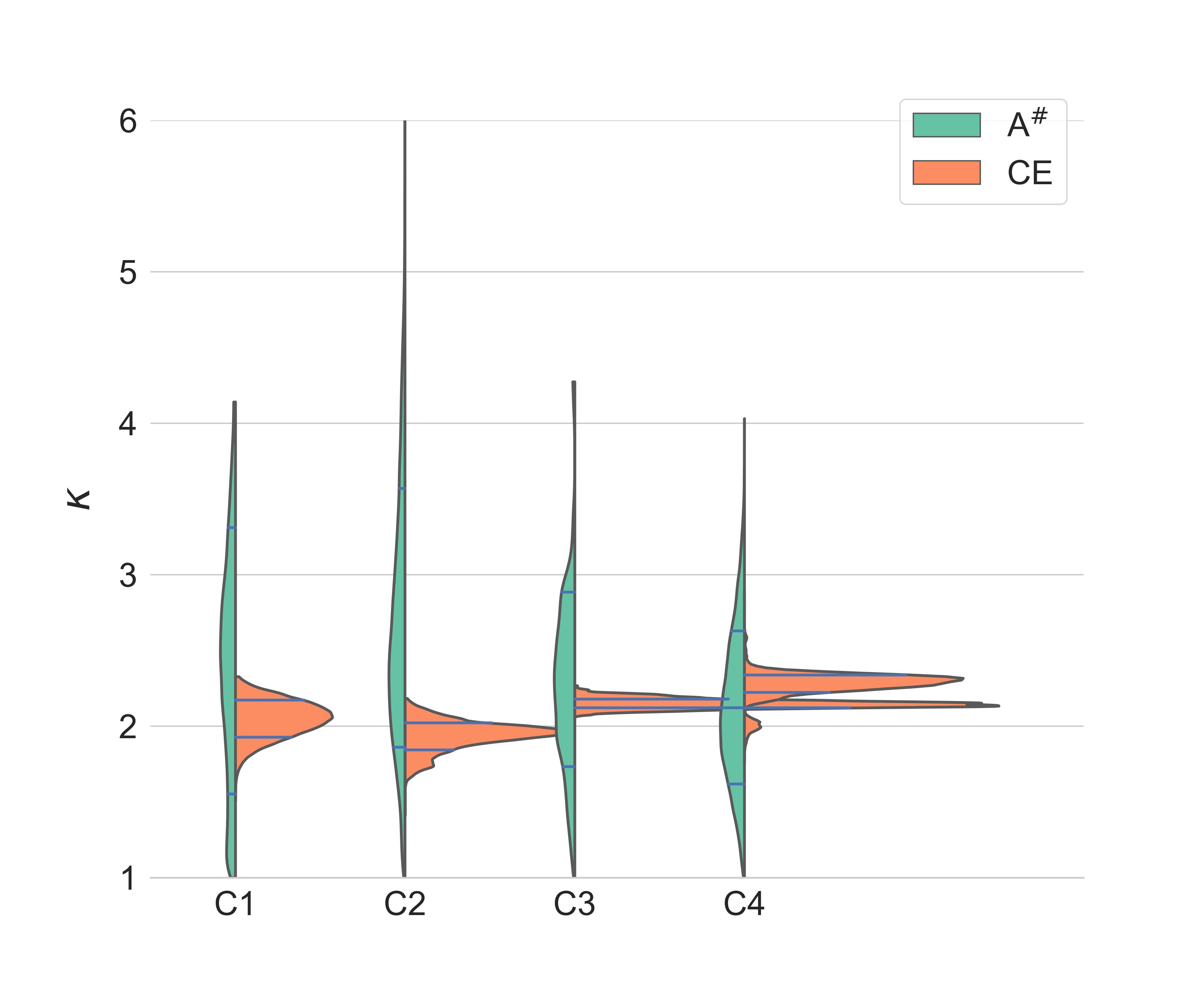}
	\caption{Left: Posterior distribution of SIQM $\kappa_2$, $\chi_1$, and $\chi_2$ with A$^\#$ and CE networks, respectively. Here the range of values above the marginalized distributions are given within a $90\%$  confidence interval. Right: Violin plot for $\kappa$ of the $2.6\, M_\odot$ object for the four cases. The violin has been split into two parts: green (left) and orange (right) present constraints from the A$^\#$ and CE networks. Each half violin has the same area and the horizontal lines show 1 $\sigma$ uncertainty bounds.}
	\label{fig:violin_corner}
\end{figure*}

\subsubsection{Results for Four Types of Sources}
In order to assess the measurement uncertainty of $\kappa$ for mass-gap objects with future GW observations, we consider different types of systems with different detector networks as shown in Tables~\ref{tab1} and \ref{tab2}. Notice that we have assumed the same distance for all systems, whereas in reality the expected distance of different types of systems may vary, depending on the underlying merger rate. For a binary at a different distance, one can easily scale the uncertainty as the measurement SNR is inversely proportional to the distance.\\

For example, with the modified IMRPhenomXPHM waveform that includes both precessing and non-precessing effects and the possible HA effect, the posterior distribution of relevant binary parameters of the C2 scenario in Table~\ref{tab2} is shown in the left panel of Fig.~\ref{fig:violin_corner}. For a GW190814-like event (like C2) with low primary spin, assuming the A$^\#$ network, the measurement uncertainty of $\kappa_2$ is around 2 with a $90\%$ upper limit placed at $\kappa \approx 3.9$. On the other hand, the CE network gives rise to a measurement of $\kappa_2 =1.95_{-0.21}^{+0.13}$, which means that the third-generation detector will be able to provide decisive evidence on the nature of the mass-gap objects, assuming they are rapidly spinning.\\

In the violin plot shown in Fig.~\ref{fig:violin_corner}, we summarize the posterior distribution of $\kappa$ of the $2.6\,M_\odot$ object from MCMC simulations corresponding to different detector networks and binary cases. It is evident that with the assumed system parameters, the A$^\#$ network is only capable of marginally constraining  the nature of the mass-gap objects, as the $\kappa=1$ case cannot be excluded with high statistical significance. While the PI mismatch of ``C3"- and ``C4"-type binaries is about two order of magnitude greater than ``C1"- and ``C2"-type binaries (as the mass ratios for C3 and C4 are much smaller than fror C1 and C2), the constraints on $\kappa$ of the $2.6 M_\odot$ object are within the same order of magnitude for all four cases. This is because in C3 and C4 both objects have $\kappa$ as a free parameter (both $\kappa_1$ and $\kappa_2$ have priors), whereas only the less massive object in C1 and C2 has an undetermined $\kappa$ (only $\kappa_2$ has a prior). 
The CE network should be able to draw decisive conclusions for all four binaries considered.\\

\section{Black hole mimickers}\label{sec:mimickers}
BH mimickers may be star-like objects (e.g. boson stars \cite{boson_1997,boson_2017}) with continuous matter/field distribution, or contain hard boundaries (e.g., gravastars \cite{gravastar_2004,gravastar_2015,gravastar_2023} and AdS bubbles \cite{bubbles_2017}) that separate out different spacetime regimes. In addition, alternative BH solutions predicted by modified theories of gravity may also be classified as BH mimickers. In order to test/constrain their existence, it is useful to examine several key observables from the gravitational wave measurements such as the tidal Love number, the SIQM, and absorption/additional dissipation effects \cite{Cardoso_2019}. Current models of boson stars could produce a $\kappa$ ranging from 10 to 150 \cite{boson_1997,Herdeiro_2014,Baumann_2019,Chia_2020}. In the future, if there are events with statistically significant $\kappa\neq 1$ from heavy objects (such that these objects cannot be NSs), it will provide  evidence for the existence of exotic objects.\\

In this section, we focus on the scope of measuring SIQMs of BH mimickers including both precessing and non-precessing effects. Since such an analysis has been performed for BBH events in the GWTC-$1$ \cite{tgr_O1}, GWTC-$2$ \cite{tgr_O2} and GWTC-$3$ \cite{GWTC3-tgr} catalogs considering only the non-precessing AI effect with precessing waveform model, we first re-perform the analysis for a few events with clear signature of spin effects, with the PI effect considered. In addition, we will discuss the prospects of such measurement for O$4$ binaries, and show that it is possible to obtain better constraints on the SIQMs.\\

One important difference for the MCMC simulations performed in this section is that instead of having  individual $\kappa$ for the two objects in the binary, we assume that $\kappa_1$=$\kappa_2$ (following the convention in \cite{GWTC3-tgr}) and define the symmetric $\kappa$ and its deviation from one:
\begin{align}
    \kappa_s = (\kappa_1+\kappa_2)/2\,,\quad \delta\kappa_s = \kappa_s - 1\,,
\end{align}
For BBHs, $\delta\kappa_s=0$. A nonzero value of $\delta\kappa_s$ with sufficient statistical significance would indicate the object is not a BH (at least, not the kind in GR).

\subsection{GWTC Events}
\begin{figure*}[ht]
	\centering
    \includegraphics[width=0.5\textwidth]{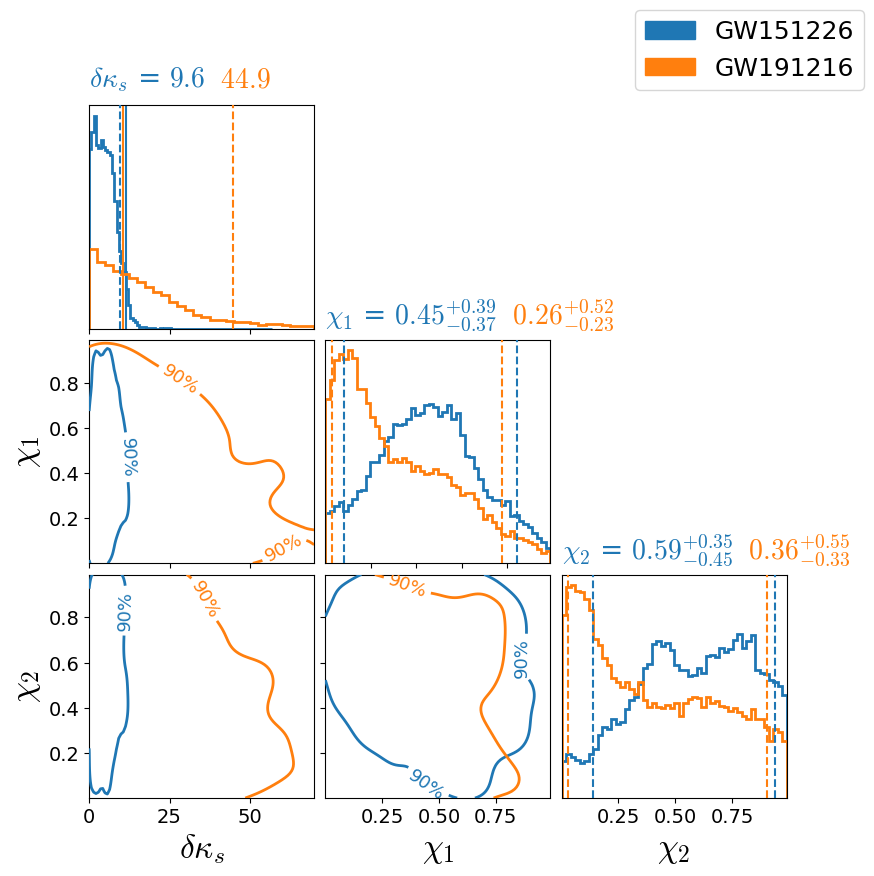}
    \includegraphics[width=0.47\textwidth]{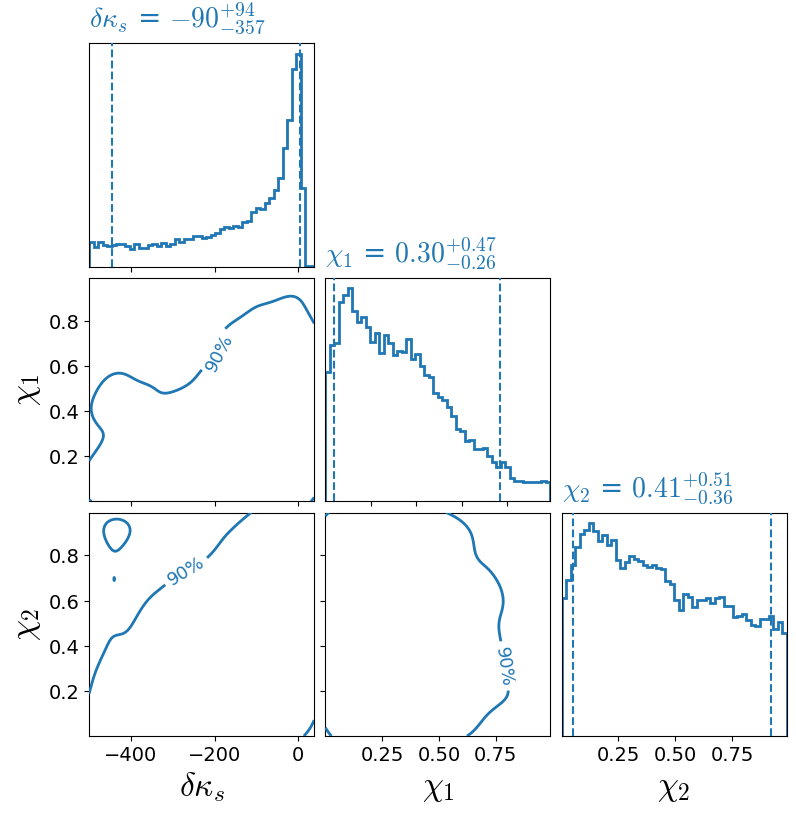}
    \caption{Posterior distribution of quadrupole moment deviations $\delta\kappa_s$, $\chi_1$ and $\chi_2$ from GW events with the IMRPhenomPv2 waveform model. Left: positive prior on $\delta\kappa_s$. The dashed lines in marginalized $\delta\kappa_s$ are the $90\%$ upper bound, while the corresponding LVK bounds are depicted by solid lines (they are around 10, very close to each other). Right: a generic prior on $\delta\kappa_s$. The result is consistent with the LVK and Krishnendu's findings \cite{tgr_O2,divyajyoti2023effect}. All other dashed lines in the marginalized distributions represent the $5\%$ and $95\%$ credible level lines.}
	\label{fig:gw_Pv2}
\end{figure*}

\begin{figure*}[ht]
	\centering
    \includegraphics[width=0.48\textwidth]{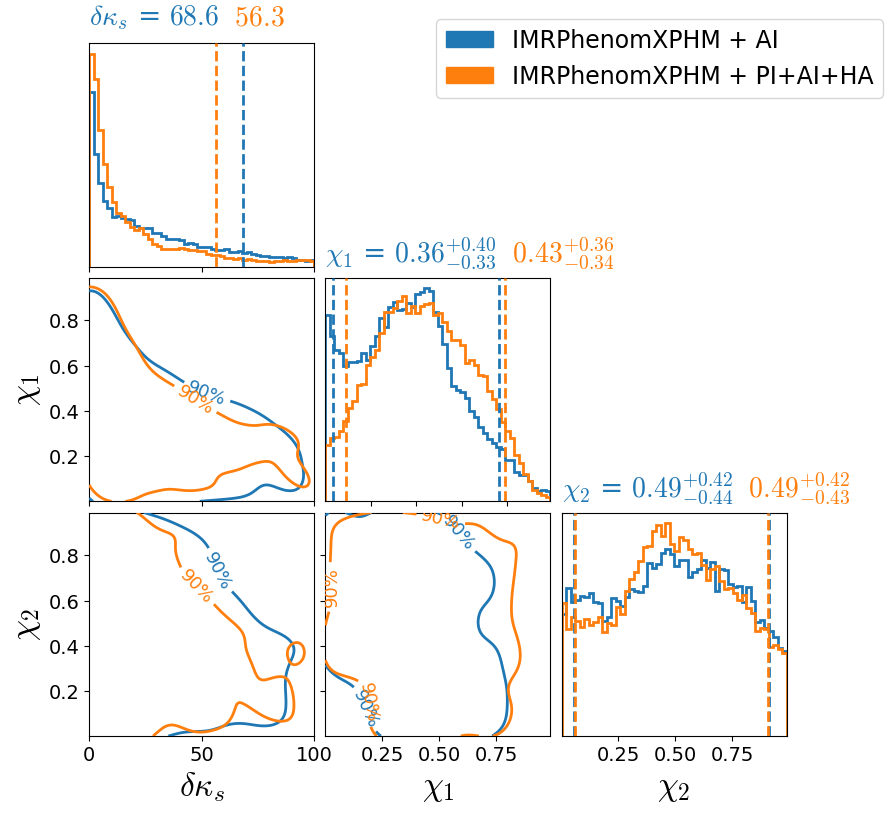}
    \includegraphics[width=0.48\textwidth]{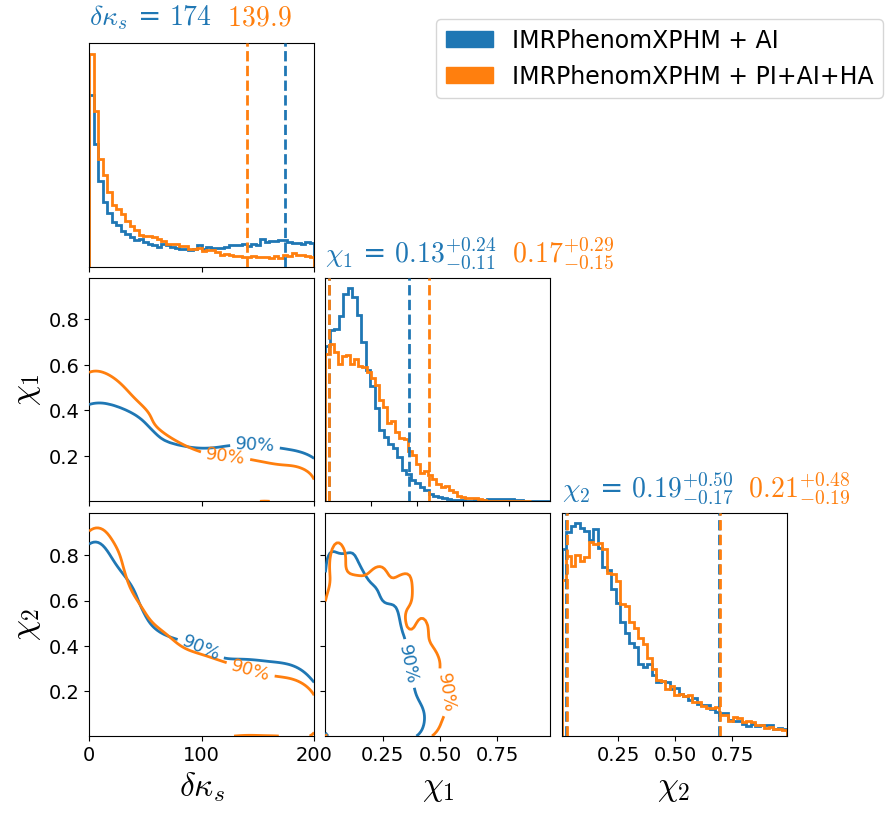}
	\caption{Posterior distribution of quadruposle moment deviations $\delta\kappa_s$, $\chi_1$ and $\chi_2$ with GW151226 (left) and GW191216\_213338 (right) with the IMRPhenomXPHM + different effect combination. The dashed lines in marginalized $\delta\kappa_s$ are $90\%$ upper bound, while they are $5\%$ and $95\%$ credible level lines for $\chi_1$ and $\chi_2$.}
	\label{fig:gw_events}
\end{figure*}

\begin{figure*}[ht]
	\centering
    \hspace{-5pt}\includegraphics[width=0.47\textwidth]{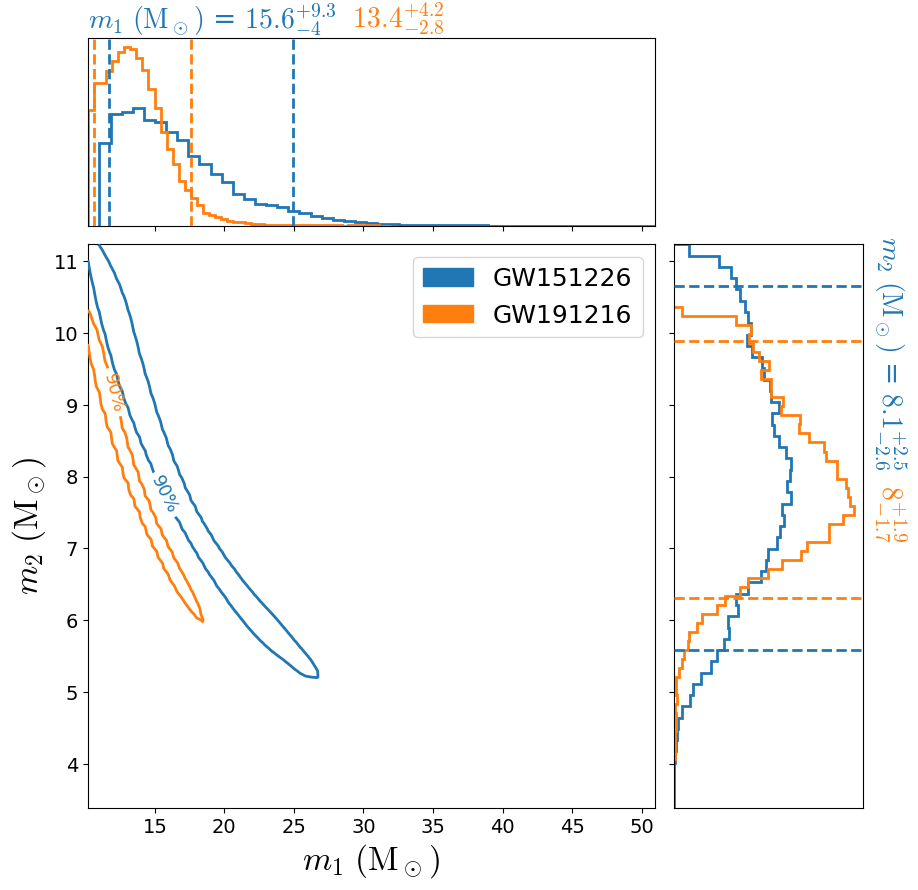}\hspace{10pt}
    \includegraphics[width=0.47\textwidth]{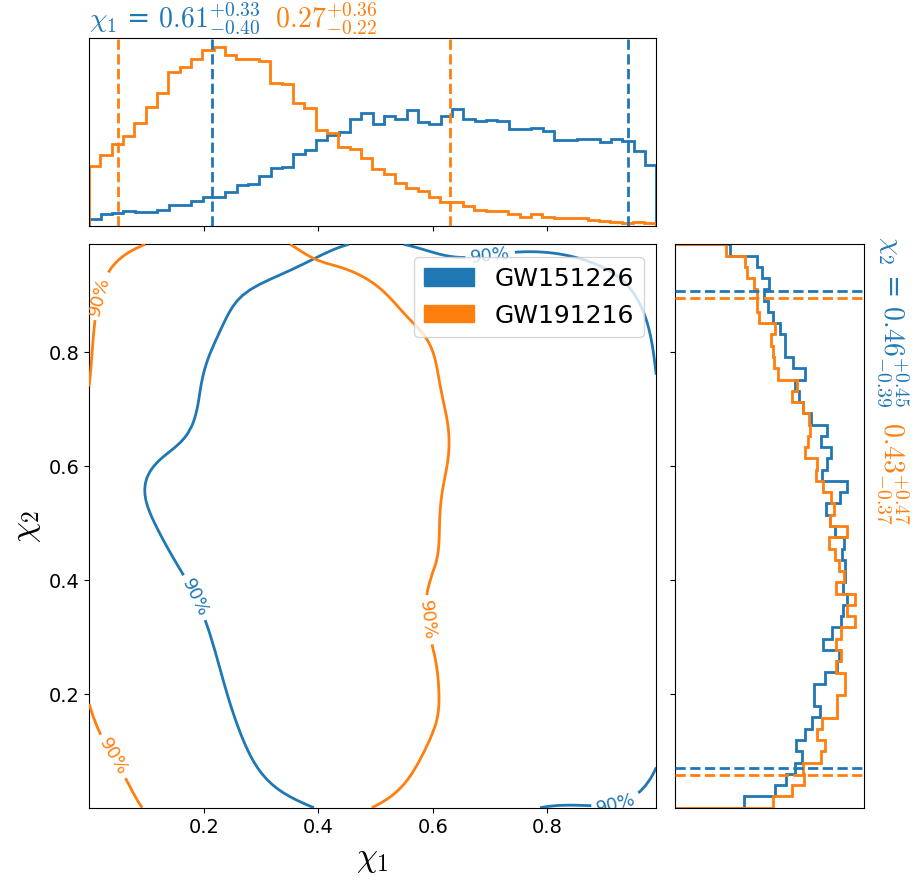}
	\caption{Posterior distribution of individual masses and spins with GW151226 and GW191216\_213338. The dashed lines in marginalized plots are $5\%$ and $95\%$ credible level lines. The posterior samples used in this plot are from public available works \cite{nitz20224ogc}.}
	\label{fig:gw_mass_spin}
\end{figure*}

Constraints of $\delta\kappa_s$ from individual GW events have been discussed using the IMRPhenomPv2 waveform model with AI effect only \cite{Khan_2019}. To further illustrate the difference between IMRPhenomPv2 and our modified waveform model (a theoretical comparison is presented in Sec.\ref{subsec:wf-model}), we choose GW events GW151226 and  GW191216\_213338 (downloaded from The Gravitational Wave Open Science Center (GWOSC) \cite{GWOSC_2023}), which show nonzero spins in the posterior and perform MCMC parameter estimation using IMRPhenomPv2 versus IMRPhenomXPHM ($l=m=2$ mode only). The resulting posterior distribution of $\delta \kappa_s$ is shown in Fig.~\ref{fig:gw_Pv2} and Fig.~\ref{fig:gw_events}, respectively. We find that the IMRPhenomPv2 waveform generally leads to much tighter constraints than the IMRPhenomXPHM waveform, which is counter-intuitive as by construction the IMRPhenomXPHM waveform describes the spin dynamics more accurately than the IMRPhenomPv2 waveform. We think that the discrepancy comes from the fact that the IMRPhenomPv2 waveform contains much larger phase error than the IMRPhenomXPHM waveform when $\delta \kappa_s$ is large, so that the large $\delta \kappa_s$ regime is more easily ruled out for the IMRPhenomPv2 waveform  (as demonstrated in the mismatch analysis in Fig.~\ref{fig:comparison-mismatch_C5}). Therefore, {\it a more accurate waveform does not always lead to tighter constraints on $\delta \kappa_s$, at least in the relatively low SNR limit.} In the high SNR limit, on the other hand, we expect that the systematic error in the IMRPhenomPv2 waveform shall lead to a more biased measurement of $\delta \kappa_s$. We will further compare these two waveforms with the injected event in the next section. Additionally, a recent study has undertaken a in-depth comparison \cite{divyajyoti2023effect}, focusing on the same two waveform models (IMRPhenomPv2 and IMRPhenomXPHM). The study has observed a similar phenomenon.

In addition, when we use the IMRPhenomPv2 waveform for parameter estimation for GW151226 and GW191216\_213338, we obtain the $90\%$ constraints on $\delta k_s$ to be $\sim 10$ for GW151226 and $\sim 40$ for GW191216\_213338. The former result is consistent with the bound listed in \cite{tgr_O2}, but the latter is about four times greater than the bound presented in \cite{GWTC3-tgr}, which is also $\sim 10$. However, as we compare the binary parameters, the component masses and the SNRs for these two events are similar, but the inferred spins of GW151226 are significantly larger than those of GW191216\_213338 (see the comparison in Fig.~\ref{fig:gw_mass_spin} and Fig.~\ref{fig:gw_polar}). Therefore, one would expect tighter constraints on $\delta \kappa_s$ for GW151226. In fact, when we use the IMRPhenomXPHM waveform for parameter estimation, we indeed also find that the constraint from GW151226 is tighter than that from GW191216\_213338 as shown in Fig.~\ref{fig:gw_events}.\\

The IMRPhenomXPHM waveform model with the PI effect included provides more accurate spin dynamics, and potentially more constraining power on $\delta \kappa_s$ because the PI effect tends to introduce additional mismatch in the waveform for precessing binaries. For GW151226 and GW191216\_213338, the resulting posterior distributions of spin parameters are shown in Fig.~\ref{fig:gw_events} in comparison with those without the PI effect. For both events, we only observe marginal improvement in constraining $\delta \kappa_s$: for GW191216\_213338 the $90\%$ confidence interval of $\delta \kappa_s$ with PI effect included is $\sim 140$ and without PI is $\sim 174$; for GW151226 the $90\%$ confidence interval of $\delta \kappa_s$ with PI effect included is $\sim 56$ and without PI is $\sim 68$. The improvement is not very significant, potentially because the spin precession in both system are still mild.

\subsection{Injected Events}
\begin{figure}[ht]
	\centering
    \includegraphics[width=0.45\textwidth]{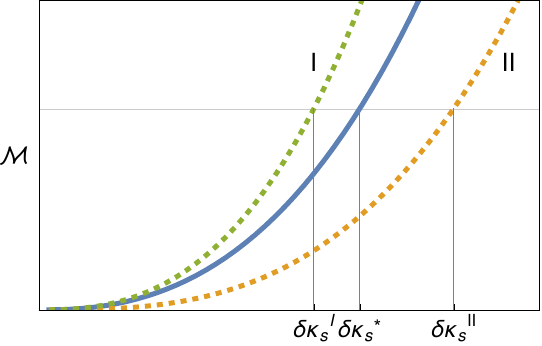}
	\caption{This cartoon plot illustrates that a more accurate waveform model does not necessarily imply better constraints on $\delta \kappa_s$ (This plot is for illustration purpose only, which is not computed out of an actual calculation. We assume the mismatches reduce to zero as $\delta \kappa_s \to 0$, in other words, they reduce to the same waveform models as $\delta \kappa_s \to 0$). Three different waveform models are shown: the true waveform in solid blue, and two alternative waveform models in dashed (green is labeled I, and yellow labeled II). For a given mismatch, the true waveform model will be able to constrain $\delta \kappa_s^\star$ to $\left[0,\delta \kappa_s\right]$. The incorrect waveform model II will give a poorer constraint, as the range $\left[0,\delta \kappa_s^{\rm II}\right]$ is larger than $\left[0,\delta \kappa_s^\star\right]$. However, the incorrect model I will give a tighter constraint.}
	\label{fig:comparison-mismatch}
\end{figure}

\begin{figure}[h]
	\centering
    \includegraphics[trim={25pt 22pt 40pt 50pt},clip,width=0.48\textwidth]{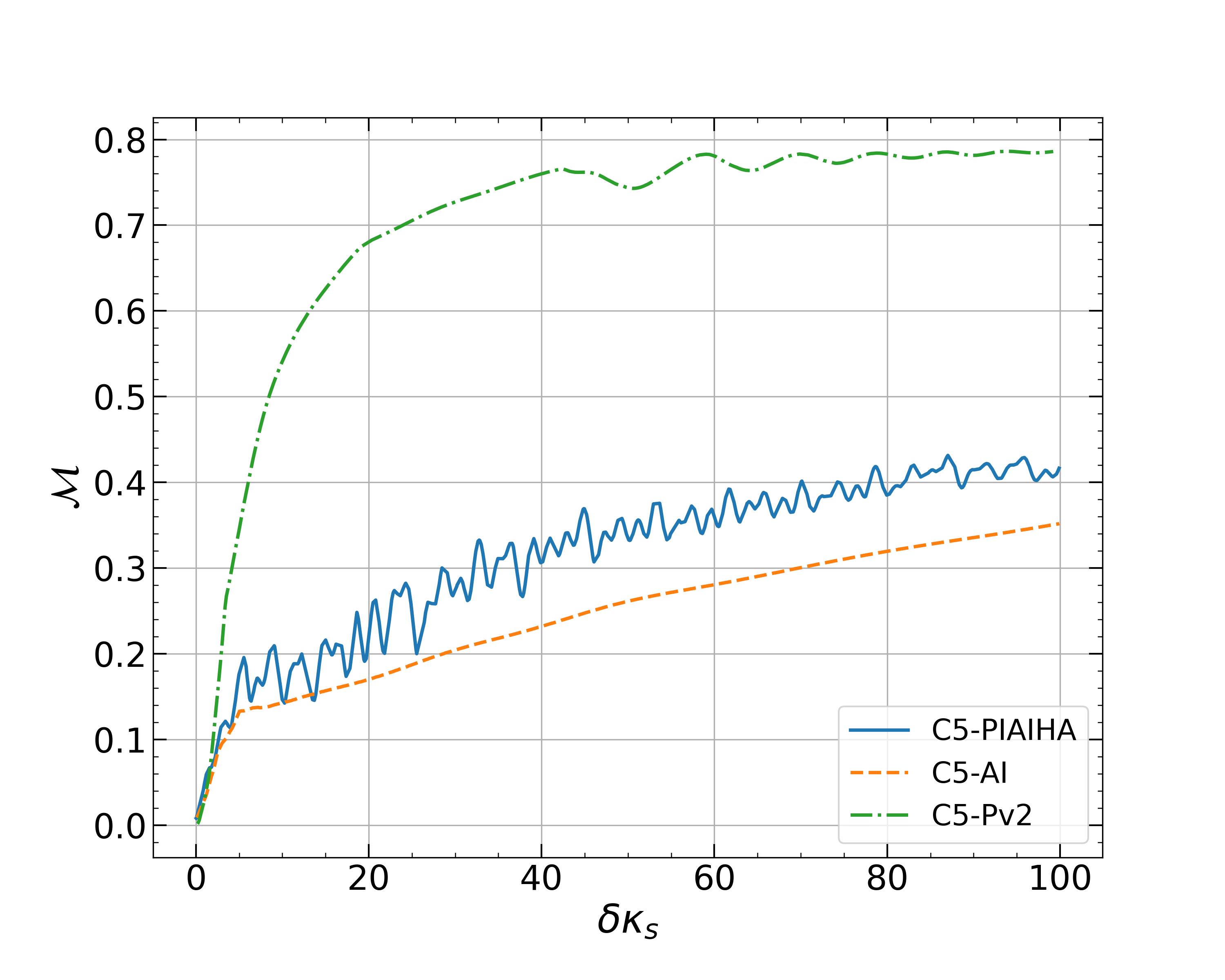}
	\caption{This plot shows the mismatch as a function of $\delta \kappa_s$ for the C5 configuration, which is calculated from actual waveform models using Eqn.~\ref{eq:mimatch2}. Three different waveform models are displayed: the most accurate waveform in blue (IMRPhenomXPHM+PI+AI+HA), along with two alternative waveform models in green (IMRPhenomPv2) and orange (IMRPhenomXPHM+AI). This demonstrates that for C5, the GW151226-like case, the IMRPhenomPv2 model will impose a more stringent constraint on $\delta \kappa_s$ for a given mismatch. This aligns with our MCMC simulation results, as depicted in Fig.~\ref{fig:gw_Pv2} and Fig.~\ref{fig:gw_events} for the GW151226 event.}
	\label{fig:comparison-mismatch_C5}
\end{figure}

\begin{figure*}[th]
	\centering
    \includegraphics[width=0.48\textwidth]{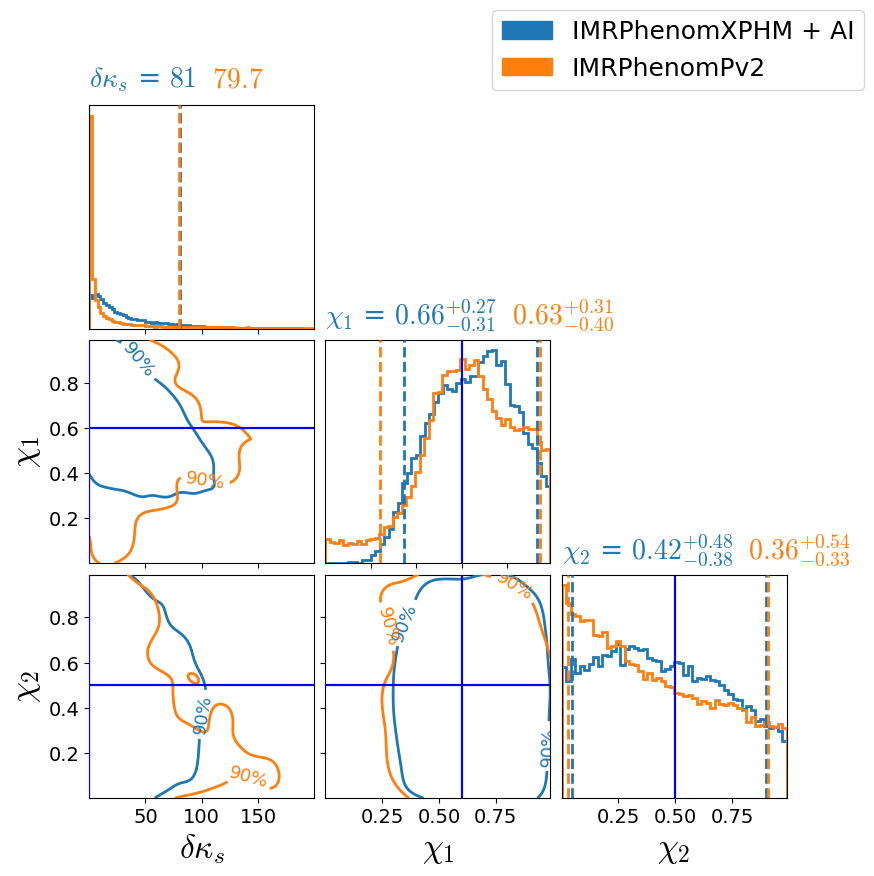}
    \includegraphics[width=0.48\textwidth]{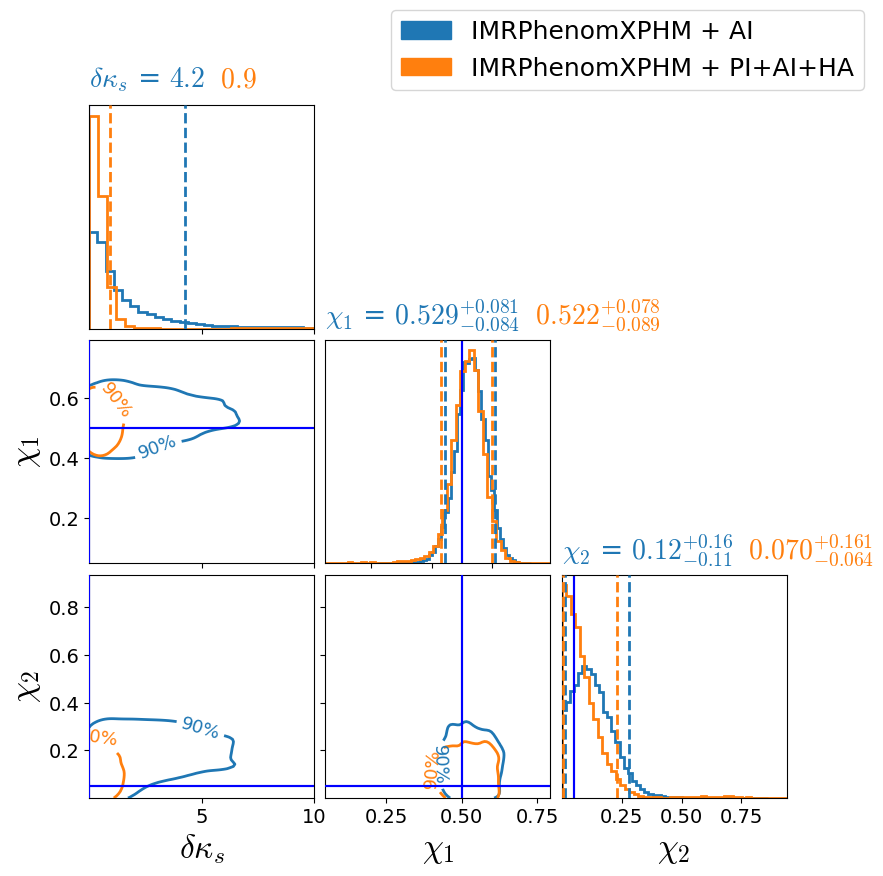}
	\caption{Posterior distribution of quadrupole moment deviations $\delta\kappa_s$, $\chi_1$ and $\chi_2$. Left: result from C5 (GW151226-like) injection with IMRPhenomXPHM+AI but recovery with IMRPhenomXPHM+AI and IMRPhenomPv2 models. Right: C6 with O4 detector network. The dashed lines in marginalized $\delta\kappa_s$ are $90\%$ upper bound, while they are $5\%$ and $95\%$ credible level lines for $\chi_1$ and $\chi_2$.}
	\label{fig:gw_mimickers}
\end{figure*}

In order to further compare the performance of the IMRPhenomPv2 waveform and the IMRPhenomXPHM waveform ($l=m=2$ mode only and without the PI effect included yet) in constraining $\delta \kappa_s$, we inject a GW151226-like event (C5, with detailed parameters shown in Table~\ref{tab3}) using the IMRPhenomXPHM waveform, and try to recover the parameters using both waveforms. The posterior distributions are shown in Fig.~\ref{fig:gw_mimickers} (left panel). We find that although the underlying event is generated by the IMRPhenomXPHM waveform, the posterior distribution  of $\delta \kappa_s$ is actually comparable if we use the IMRPhenomPv2 waveform for recovery. The reason is that the IMRPhenomPv2 waveform produces much larger mismatch for large $\delta \kappa_s$, so it appears to be more constraining than the more accurate waveform model. 
In general, as we compute the mismatch between the $\delta \kappa_s=0$ waveform and a $\delta \kappa_s \neq 0$ waveform as shown in Eq.~\ref{eq:mimatch2} and illustrated in a cartoon plot in Fig.~\ref{fig:comparison-mismatch}, the true waveform model should have its mismatch depend on $\delta \kappa_s$ with certain function dependence, while the inaccurate waveform model may have its mismatch larger (model I in Fig.~\ref{fig:comparison-mismatch}) or smaller (model II) than the  mismatch of the true waveform for the same $\delta \kappa_s$. If the mismatch of the inaccurate waveform model is larger than the true mismatch, it  appears to be more constraining than the true model, and vice versa. In other words, when they have the same mismatch values, model I  gives a tighter constraint on $\delta \kappa_s$,
\begin{equation}\label{eq:mimatch2}
\mathcal{M}(\delta \kappa_s) = 1 - \max_{t_c,\phi_c}\, \frac{<h(0), h(\delta \kappa_s)>}{\sqrt{<h(0), h(0)>}\,\sqrt{<h(\delta \kappa_s), h(\delta \kappa_s)>}}\,.
\end{equation}
where $h(0)=h(\delta \kappa_s = 0)$. $\mathcal{M}(\delta \kappa_s)\rightarrow 0$ while $\delta\kappa_s \rightarrow 0$.

We can also use injected events to compare the performance of IMRPhenomXPHM waveforms with or without the PI effect included.
To do this, we construct an O4-type network (LIGO Hanford, LIGO Livingston and Virgo) to test one case of BH mimickers (C6) .  The binary configuration we assume is shown in Table~\ref{tab3}, with one of the BHs having a significant spin magnitude $\chi_1=0.5$, and the spin of the other BH is small $\chi_2 =0.05$.  With the assumed distance at $200$ Mpc, the network SNR is approximately 110. Notice that if a similar system is detected at a different distance, one can scale the uncertainties accordingly, as the SNR is inversely proportional to the distance.  For the injected waveform, we also assume that these are two BHs with $\kappa_1=\kappa_2=1$, hence $\delta\kappa_s=0$. Among the randomly generated spin orientation and inclination angles, we pick a set of parameters such that the mismatch contribution from the PI and AI effect are also roughly $50\% : 50\%$, so that we can assess the constraining power of the PI effect relative to the AI effect, since based on mismatch alone they should contribute evenly.\\

In the Fig.~\ref{fig:gw_mimickers} (right panel), we  present the posterior distribution of $\delta\kappa_s$ and the dimensionless spin magnitudes. With the PI effect included, there is a much tighter constraint on the magnitude of $\delta \kappa_s$. In addition, the measurement uncertainty on $\chi_2$ (the spin of the slowly-rotating BH) also becomes significantly better than the distribution recovered without using the PI effect. Although we have not made further comparison using more simulated data in different scenarios, it is reasonable to expect that the waveform model with the PI effect included will improve the measurement of $\delta \kappa_s$ for at least some of the O4 events.
Even if it did not offer significant additional constraining power, physically the PI effect is present if $\delta \kappa_s$ is nonzero and the system is precessing.  To produce results as accurately as possible and without bias, we should apply the more complete waveform model, since there is little additional computational overhead associated in doing so \cite{Pratten_2020imr3}.

\renewcommand{\arraystretch}{1.5}
\begin{table*}[th]
\centering
\begin{tabularx}{0.99\textwidth} {>{\centering\arraybackslash}X |>{\centering\arraybackslash}X | >{\centering\arraybackslash}X | >{\centering\arraybackslash}X | >{\centering\arraybackslash}X | >{\centering\arraybackslash}X | >{\centering\arraybackslash}X | >{\centering\arraybackslash}X | >{\centering\arraybackslash}X | >{\centering\arraybackslash}X | >{\centering\arraybackslash}X | >{\centering\arraybackslash}X | >{\centering\arraybackslash}X | >{\centering\arraybackslash}X | >{\centering\arraybackslash}X | >{\centering\arraybackslash}X}
\hline
\hline
\multicolumn{2}{c|}{} & \multicolumn{2}{|c|}{$(m_1, m_2) (M_\odot)$} & $(\kappa_1, \kappa_2)$ & $(\chi_1, \chi_2)$ & \multicolumn{2}{c|}{$(\theta_1, \theta_2)$} & \multicolumn{2}{c|}{$(\phi_1, \phi_2)$} & $\iota$ & $(\alpha, \delta)$ & $d_\mathrm{L}$(Mpc) & \multicolumn{2}{c}{SNR}\\
\hline
\multicolumn{2}{c|}{\;\; C5 \;\; } & \multicolumn{2}{|c|}{(16, 8)} & (1, 1) & (0.6, 0.5) & \multicolumn{2}{c|}{(1.0, 1.4)} & \multicolumn{2}{c|}{(3.2, 3.1)} & 1.3 & (3.5, -0.2) & 490 & \multicolumn{2}{c}{15.5(aLIGO)} \\
\multicolumn{2}{c|}{ C6 } & \multicolumn{2}{|c|}{(30, 20)} & (1, 1) & (0.5, 0.05) & \multicolumn{2}{c|}{(1.02, 0.5)} & \multicolumn{2}{c|}{(6.26, 2.49)} & 2.46 & (1, 2) & 200 & \multicolumn{2}{c}{110(O4), 443((A$^\#$)} \\
\hline
\hline
\end{tabularx}
\caption{Configurations of two BBH injections used for MCMC simulations, where C5 is a GW151226-like injection. The last column shows SNRs for the GW151226-like event, which is close to the real event SNR, and for C6 with O4 network and A$^\#$ sensitivity, respectively.}
\label{tab3}
\end{table*}

\section{Conclusions}
We have examined the effect of the SIQM on the resulting waveform, and the constraints that can be placed on the SIQM of objects as a result.  We have shown that, at least for some precessing configurations, by including the effect of the SIQM on the precession description of spins and angular momentum, it is possible to achieve better measurement accuracy on $\kappa$.
This has important consequences in assessing the nature of mass gap objects where other methods, namely the tidal deformability, HA (or lack thereof), and EM counterpart, may fail to distinguish between the heavy NSs and light BHs.\\

In our analysis, we used the waveform model that includes the complete SIQM effect \cite{LaHaye_2022} and run MCMC simulations. We find that for another GW190814-like event, with an A$^\#$ type detector network, the statistical significance for individual events is generally not sufficient to distinguish between the nature of objects with high confidence. However, for a CE network we should easily be able to distinguish between  NSs and BHs for such mass-gap objects.\\

Second, we find that, using this waveform model, even with O4 sensitivity we can place tighter constraints than similar waveform models that ignore this effect.  This is a result of the PI effect influencing the constraining power more than the aligned-spin effect.  This, of course, with the caveat that the system must be exhibiting significant precession. On the other hand, through the illustration with injected events and real events from the GWTC catalog, we have shown that a more accurate waveform does not necessarily lead to tighter constraints on $\delta \kappa_s$ for the tests of BH mimickers. Fortunately, with the state-of-the-art IMRPhenomXPHM waveform, it appears that by including the PI effect the waveform becomes both more accurate and more constraining in determining $\delta \kappa_s$.\\

To reiterate on our previous work, one potential avenue of improvement is improving the computational efficiency by producing a more analytic solution to the spin dynamics.  Another potential addition to the waveform in the future is the inclusion of eccentricity.  Since eccentricity and precession can produce similar effects on the waveform, not including eccentricity in the waveform model can produce posteriors that entirely exclude the true system parameters.\\

\acknowledgements
H. Y. is supported by the Natural Sciences and
Engineering Research Council of Canada and in part by
Perimeter Institute for Theoretical Physics.
Research at Perimeter Institute is supported in part by the Government of Canada through the Department of Innovation, Science and Economic Development Canada and by the Province of Ontario through the Ministry of Colleges and Universities. 
This work is supported by the National Natural Science Foundation of China (No. 11975027). 
This material is based upon work supported by NSF's LIGO Laboratory which is a major facility fully funded by the National Science Foundation.

This research has made use of data or software obtained from the Gravitational Wave Open Science Center (gwosc.org), a service of the LIGO Scientific Collaboration, the Virgo Collaboration, and KAGRA. This material is based upon work supported by NSF's LIGO Laboratory which is a major facility fully funded by the National Science Foundation, as well as the Science and Technology Facilities Council (STFC) of the United Kingdom, the Max-Planck-Society (MPS), and the State of Niedersachsen/Germany for support of the construction of Advanced LIGO and construction and operation of the GEO600 detector. Additional support for Advanced LIGO was provided by the Australian Research Council. Virgo is funded, through the European Gravitational Observatory (EGO), by the French Centre National de Recherche Scientifique (CNRS), the Italian Istituto Nazionale di Fisica Nucleare (INFN) and the Dutch Nikhef, with contributions by institutions from Belgium, Germany, Greece, Hungary, Ireland, Japan, Monaco, Poland, Portugal, Spain. KAGRA is supported by Ministry of Education, Culture, Sports, Science and Technology (MEXT), Japan Society for the Promotion of Science (JSPS) in Japan; National Research Foundation (NRF) and Ministry of Science and ICT (MSIT) in Korea; Academia Sinica (AS) and National Science and Technology Council (NSTC) in Taiwan.

\appendix
\section{Posterior distributions of spin polar angles of GW151226 and GW191216\_213338}

\begin{figure}[ht]
	\centering    \includegraphics[width=0.47\textwidth]{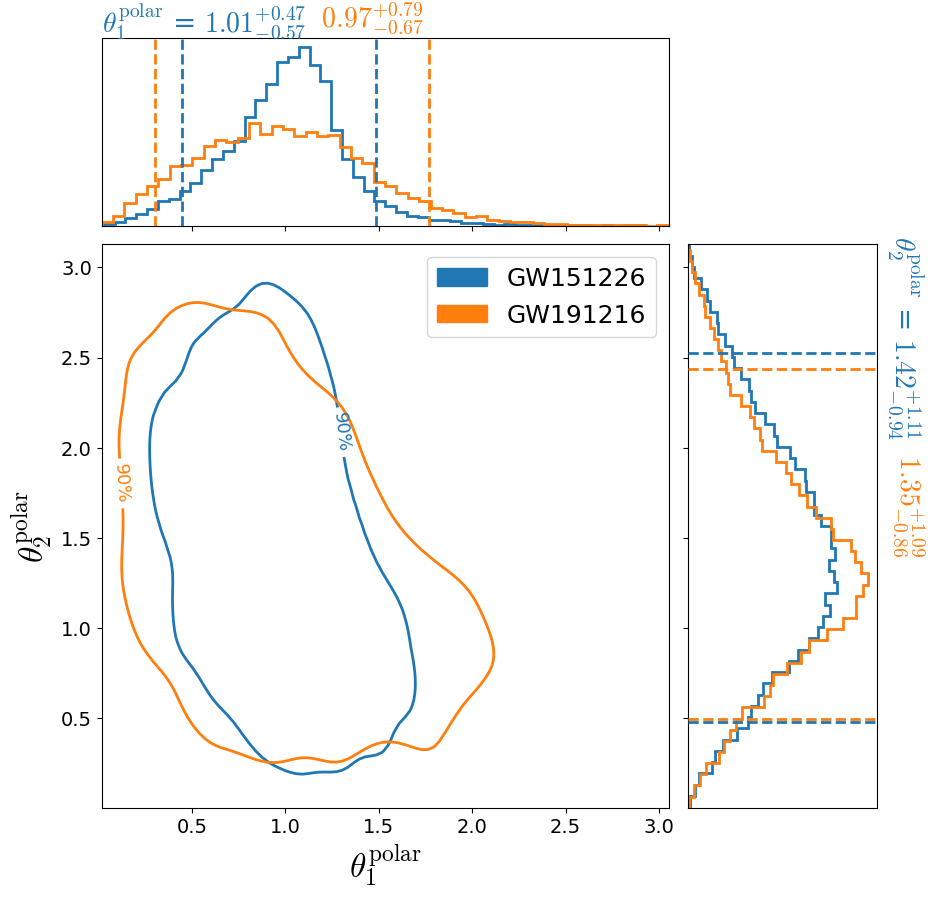}
	\caption{Posterior distribution of individual spin polar angles with GW151226 and GW191216\_213338. The dashed lines in marginalized plots are $5\%$ and $95\%$ credible level lines. The posterior samples used in this plot are from public available works \cite{nitz20224ogc}.}
	\label{fig:gw_polar}
\end{figure}

\bibliography{ref.bib}

\end{document}